\documentclass[showpacs,aps,prc,preprint,nofootinbib,superscriptaddress]{revtex4}
\usepackage[utf8]{inputenc}
\usepackage{graphicx,subfigure}
\usepackage{float}
\usepackage{amsmath}
\usepackage{amssymb}
\usepackage{multirow}
\def\bm{\boldsymbol}
\newcommand{\bea}{\begin{eqnarray}}
\newcommand{\eea}{\end{eqnarray}}
\newcommand{\be}{\begin{eqnarray}}
\newcommand{\ee}{\end{eqnarray}}
\newcommand{\no}{\nonumber \\}

\def\vp{{\bm p}}
\def\vq{{\bm q}}
\def\vk{{\bm k}}

\def\vx{{\bm x}}
\def\vy{{\bm y}}

\def\vr{{\bm r}}
\def\vs{{\bm\sigma}}
\def\la{\langle}
\def\ra{\rangle}

\begin{document}

%\preprint{}

\title{Parity violation in radiative neutron capture on deuteron}


\author{Young-Ho Song}
\email[]{song25@mailbox.sc.edu}
\affiliation{Department of Physics and Astronomy, University of South Carolina, Columbia, SC, 29208}
%\thanks{}

\author{Rimantas Lazauskas}
\email[]{rimantas.lazauskas@ires.in2p3.fr}
\affiliation{IPHC, IN2P3-CNRS/Universit\'e Louis Pasteur BP 28,
F-67037 Strasbourg Cedex 2, France}

\author{Vladimir Gudkov}
\email[]{gudkov@sc.edu}
\affiliation{Department of Physics and Astronomy, University of South Carolina, Columbia, SC, 29208}

%\thanks{}

%\thanks{}


\date{\today}

\begin{abstract}
Parity violating (PV) effects in neutron-deuteron radiative capture
are studied using Desplanques, Donoghue, and Holstein (DDH)
and effective field theory weak potentials.
The values of PV effects are calculated using wave functions,
obtained by solving three-body Faddeev equations in configuration space for phenomenological strong potentials.
The relations between physical observables and low-energy constants are presented, and dependencies of the calculated PV effects on
strong and weak potentials are discussed.
The presented analysis shows the possible reason for the existing discrepancy in PV nuclear data analysis using the DDH approach and
reveals a new opportunity to study  short range interactions  in nuclei.
\end{abstract}

%\keywords{}

\maketitle

\section{Introduction}
Low energy parity violating (PV) effects  play  an important role in understanding the main features of the Standard model.
Many nuclear PV effects were measured and calculated during the last several years.
Despite the fact that existing  calculations of nuclear PV effects are
in a reasonably good agreement with the measured ones,
lately it became clear (see, for example \cite{Zhu:2004vw,HolsteinUSC,DesplanqueUSC,RamseyMusolf:2006dz}  and references therein)
that it is rather difficult to describe the available experimental data
with the same set of weak nucleon coupling constants
using the traditional
DDH \cite{Desplanques1980} weak meson exchange potential.

As a possible solution for this problem, a  new approach, based on the effective field theory (EFT), has been introduced
to parameterize the PV effects in a model independent way (see, papers \cite{Zhu:2004vw,RamseyMusolf:2006dz,Liu:2006dm}
and references therein).
The main goal of the EFT approach is to describe a large number of PV effects in terms of a small number of constants (LEC),
 which are the free parameters of the theory.
Unfortunately, since the number of experimentally measured
(and independent in terms of unknown LECs)
PV effects in two body systems is not enough to constrain all LECs~\cite{Girlanda:2008ts,Phillips:2008hn,Shin:2009hi,Schindler:2009wd}.
In order to determine these constants it is  necessary to include also the data obtained on heavier nuclear systems.

 Furthermore one should better understand  PV effects  in heavier nuclei because
 these effects might be essentially enhanced \cite{Sushkov:1982fa,Bunakov:1982is,Gudkov:1991qg} in many body systems.
However,  how to apply the EFT approach for the calculations of PV effects in nuclei it is still an open question.

To verify the possible issues related to the application of the DDH description
of PV effects in nuclei and the possibility of  systematic calculations of PV effects in nuclei using EFT approach, it is desirable to start from the
 calculations of PV effects in the simplest nuclear systems, such as  neutron-deuteron (n-d) compound.
PV effects for the elastic n-d scattering have been calculated recently \cite{Schiavilla:2008ic,Song:2010sz} using both
DDH and EFT approaches. However, before extending these techniques to the many-body nuclear systems,
 it is  important to consider inelastic processes which are usually more sensitive to  short range interactions.

With this aim, we present in this paper  a comprehensive analysis of PV effects in neutron-deuteron radiative capture \cite{Moskalev:1969,Hadjimichael:1974,McKellar:1974yr,Desplanques:1986cq}
using weak potential of DDH-type, as well as weak potentials obtained in pionless and pionful EFT.
For the strong interaction, we have tested several realistic nucleon-nucleon potentials, also in conjunction with  three-nucleon force.
Three-nucleon wave functions have been obtained by solving Faddeev equations in configuration space
for the complete Hamiltonians comprising both weak and strong interactions.

The paper is structured as follows. In the next section, a brief description of the employed formalism is presented.
Then, we discuss the results of our calculations and perform a detailed analysis  of
model and  cutoff dependence of the calculated PV parameters.
In conclusion, the implications of our result are summarized.

\section{Formalism}

We consider three parity violating observables in the radiative
neutron capture on deuterons ($n+d\to ^3H+\gamma$):
circular polarization of the  emitted photons ($P^\gamma$),
 asymmetry of the photons in relation to neutron polarization ($a_n^\gamma$),
and asymmetry of the photons in relation to deuteron polarization ($A_d^\gamma$).
For  low energy neutrons, the expressions for these PV effects could be written in terms of  parity conserving magnetic dipole ($M1$)
%transition matrix elements
and parity violating electric dipole ($E1$) transition
matrix elements as:
\bea
a^\gamma_n(E)&=&
   \frac{2}{3}\frac{ {\rm Re} \left[
   \sqrt{2}(E1^{*}_{\frac{3}{2}} M1_{\frac{1}{2}}
        +E1^{*}_{\frac{1}{2}} M1_{\frac{3}{2}})
   +\frac{5}{2}(E1^{*}_{\frac{3}{2}} M1_{\frac{3}{2}})
   -(E1^{*,(+)}_{\frac{1}{2}}M1_{\frac{1}{2}})
   \right]}
        {|M1_{\frac{1}{2}}|^2+|M1_{\frac{3}{2}}|^2},\no
P^\gamma(E)&=&\frac{2
{\rm Re}\left[E1^*_{\frac{1}{2}}M1_{\frac{1}{2}}+E1^*_{\frac{3}{2}}M1_{\frac{3}{2}}\right]
}{|M1_{\frac{1}{2}}|^2+|M1_{\frac{3}{2}}|^2},\no
A^\gamma_d(E)&=&
-\frac{{\rm Re}\left[-5 E1^*_{\frac{3}{2}} M1_{\frac{3}{2}}
                     -4 E1^*_{\frac{1}{2}} M1_{\frac{1}{2}}
                    +\sqrt{2}E1^*_{\frac{3}{2}} M1_{\frac{1}{2}}
                    +\sqrt{2}E1^*_{\frac{1}{2}} M1_{\frac{3}{2}}
\right]}{2(|M1_{\frac{1}{2}}|^2+|M1_{\frac{3}{2}}|^2)}.
\eea
Here  the M1 and E1 amplitudes are defined as reduced matrix elements of
the multipole operators
\bea
X1_J\equiv \la -\vq, J_B||\hat{T}_1^X|| J\ra,\quad
\mbox{ with } X=(M,E),
\eea
where $J_B$ and $J$ are total angular momenta of bound state and scattering
state respectively, and $\vq$ is a momentum of the outgoing photon.
The electromagnetic multipole operators
in the limit of small $q$ can be written as
\bea
{\hat T}^{Mag}_{JM}(q)&\simeq&-\frac{q^J}{i(2J+1)!!}\sqrt{\frac{J+1}{J}} \int d{\bf x}
           [\hat{\bf \mu}({\bf x})+\frac{1}{J+1}{\bf r}\times \hat{\bf J}_c(x)]\cdot\nabla(x^J Y_{JM})\no
{\hat T}^{El}_{JM}(q)&\simeq&\frac{q^J}{(2J+1)!!}\sqrt{\frac{J+1}{J}}  \int d{\bf x}(
               x^J Y_{JM}\hat{\rho}(x)-\frac{iq}{J+1}\hat{\bf \mu}(x)\cdot[\vr\times\nabla x^J Y_{JM}])\nonumber ,
\eea
where $\hat{J}_c(x)$ is a convection current, $\hat{\mu}(x)$ is a magnetization
current, $\hat{\rho}(x)$ is a charge operator, and
$q=\omega$ is the energy of photon.
In our calculations, we use $M1$ operator up to $N^3LO$ in chiral order counting,
which includes contributions from two-pion exchange and contact currents obtained
in heavy baryon chiral perturbation theory\cite{Song:2008zf}.
For calculations of E1 amplitudes at the leading order, we use only E1 charge operator, which is related to 3-vector currents by Siegert's theorem.
Since,  in the used spherical harmonics convention both
parity conserving $M1$ and
parity violating $E1$ amplitudes
are purely imaginary, it is convenient to define real-valued
$\widetilde{\cal M}_J$ and
$\widetilde{\cal E}_J$ matrix elements as
\bea
M1_{J}&=&i\frac{\omega\mu_N}{\sqrt{6\pi}\sqrt{4\pi}}\widetilde{\cal M}_{J},
\quad
E1_J=-i\frac{\omega}{\sqrt{6\pi}}\widetilde{\cal E}_J,
\eea
 where $\mu_N=\frac{1}{2m_N}$.

The calculations of parity conserving $M1$ amplitudes for radiative n-d capture  have been  reported
 in papers \cite{Song:2008zf,Pastore:2009is} using the hybrid method, where the
wave functions were obtained from phenomenological potential models
and the current operators were derived from the heavy baryon chiral effective field theory.
The results of these calculations
can be approximated \cite{Song:2008zf} by the following expressions
\footnote{The sign convention of $M1$ changed from
\cite{Song:2008zf} calculation to be consistent with the convention used in this work.}
\bea
\widetilde{\cal M}_{\frac{1}{2}}
&=&+21.87+10.76[(B_{model}/B_{exp})^{-2.5}-1] \mbox{ fm}^{\frac{3}{2}},\no
\widetilde{\cal M}_{\frac{3}{2}}
&=&-12.24-11.35[(B_{model}/B_{exp})^{-2.5}-1]\mbox{ fm}^{\frac{3}{2}} \label{EQ_M1},
\eea
where two low energy constants  of
two-body M1 operators are fixed \cite{Song:2008zf} by experimental values
 of  $^3H$ and $^3He$  magnetic moments. In these expressions,  M1 amplitudes and the  binding energy of $^3H$, $B_{model}$  depends on the strong interaction model.   However, the observed explicit
correlation between the calculated
The M1 amplitudes and the  binding energy $B_{model}$ provides the unique opportunity to eliminate the model dependence. This might be done
by setting in eq.(\ref{EQ_M1})   $B_{model}/B_{exp}=1$.
M1 amplitudes obtained in such a way lead to the value of the total neutron-deutron radiative capture cross section
$\sigma_{tot}=0.49(1)$ mb, which is well consistent with the experimental data.

E1 amplitudes are calculated using three-body wave functions, which are obtained by solving Faddeev equations
in configuration space. We have tested different combinations of strong and weak potentials.
For the strong (parity conserving) part of the Hamiltonian,
we choose one of the realistic nucleon-nucleon interaction models, namely:  AV18,
 Reid, NijmII and INOY were employed. Also we have performed calculations for AV18 NN potential
 in conjunction with UIX three-nucleon force (denoted as AV18+UIX).
For the parity violating part of the Hamiltonian - one of the  weak potentials was employed, which was treated as
perturbation. In this paper, we consider three types of parity violating weak potentials:
 the standard DDH potential with meson exchange
 nucleon-nucleon interactions, the  potential derived  from
pionless  version, and the potential
 derived from  pionful version of effective field theory.
  Our approach could be considered as a hybrid method, which is similar to the hybrid approach in the line of
Weinberg's scheme and which has been successfully applied for the calculations of
 weak and electromagnetic processes involving
three-body and four-body hadronic systems \cite{Song:2007bj,Song:2008zf,Lazauskas:2009nw,
Park:2002yp,Pastore:2009is,Girlanda:2010vm}, as well as for calculations of parity violating \cite{Song:2010sz} and
time reversal violating effects in elastic
n-d scattering \cite{Song:2011sw,Song:2011jh}.
It is worth  mentioning
that alternative calculations of parity
violating effects  in elastic
n-d scattering using pionless EFT  \cite{Griesshammer:2011md} are well consistent
with the hybrid calculations \cite{Song:2010sz}, though
the detailed comparison between these two methods is required.

\subsection{The parity violating potentials}
To understand the possible difference in the description of
parity violating effects by DDH and EFT-type of potentials,
we  compare the operator structure of the potentials for the DDH potential \cite{Desplanques:1979hn}
and for two different choices of EFT potentials \cite{Zhu:2004vw} which are
 derived from pionless  and   pionful EFT Lagrangian.
 All these potentials can be expanded in terms of $O^{(n)}_{ij}$ operators \cite{Schiavilla:2008ic}  as
\bea
v_{ij}^\alpha=\sum_{n} c_n^\alpha O^{(n)}_{ij},\quad
\mbox{$\alpha=$ DDH,  pionless EFT or pionful EFT}
\eea
with the explicit forms for the operators $O^{(n)}_{ij}$ and corresponding  parameters $c_n^\alpha$
listed in  table \ref{tbl:pvpotential}
\footnote{
Note that we changed the relation between
the coefficient $C_6^{\not{\pi}}$ in the weak Lagrangian\cite{Schiavilla:2008ic}
and the coefficient $c_1^{\not{\pi}}$ of weak potential
from that of previous paper\cite{Song:2010sz},
$c_1^{\not{\pi}}=\frac{2\mu^2}{\Lambda^3_\chi}   C^{\not{\pi}}_6$,
because of the inconsistency in the convention.
However, it does not affect our results in \cite{Song:2010sz} because
we calculated matrix elements of the operators $O^{(1)}_{ij}$.      
},
%\footnote{For the standard DDH approach, the operator 12 is related to the mixing of $\rho$ and $\omega$ for
%$f_\rho(r)=f_\omega(r)=\frac{e^{-m_\rho r}}{4\pi r}$.}
where
coefficients $c_n^\alpha$ have dimension of $[\mbox{fm}]$
and scalar functions $f_n^\alpha(r)$ have
dimension of $[\mbox{fm}^{-1}]$.

\begin{table}%[H] add [H] placement to break table across pages
\caption{\label{tbl:pvpotential}
 Parameters and operators of parity violating potentials.
$g_A=1.26$, $F_\pi=92.4$ MeV.
${\cal T}_{ij}\equiv (3\tau_i^z\tau_j^z-\tau_i\cdot\tau_j)$.
Scalar function
$\tilde{L}_\Lambda(r)\equiv 3L_\Lambda(r)-H_\Lambda(r) $.
}
\begin{ruledtabular}
\begin{tabular}{cccccccc}
% Lines of table here ending with \\
$n$ & $c_n^{DDH}$ & $f_n^{DDH}(r)$ & $c_n^{\not{\pi}}$ & $f_n^{\not{\pi}}(r)$ & $c_n^{\pi}$ & $f_n^{\pi}(r)$ & $O^{(n)}_{ij}$ \\
\hline
$1$ & $+\frac{g_\pi }{2\sqrt{2} m_N}h_\pi^1$ & $f_\pi(r)$ &
      $-\frac{\mu^2 C_6^{\not{\pi}}}{\Lambda_\chi^3}$  &
      $f^{\not{\pi}}_\mu(r)$ &
      $+\frac{g_\pi }{2\sqrt{2} m_N}h_\pi^1$ & $f^\pi_\Lambda(r)$ &
      $(\tau_i\times\tau_j)^z(\vs_i+\vs_j)\cdot{\bm X}^{(1)}_{ij,-}$
\\
$2 $ & $ -\frac{g_\rho}{m_N}h_\rho^0 $ & $ f_\rho(r) $ & $
      0 $ & $ 0 $ & $
      0 $ & $ 0 $ & $
      (\tau_i\cdot\tau_j)(\vs_i-\vs_j)\cdot{\bm X}^{(2)}_{ij,+}$
\\
$3 $ & $ -\frac{g_\rho(1+\kappa_\rho)}{m_N} h_\rho^0 $ & $f_\rho(r) $ & $
      0 $ & $ 0$ & $
       0 $ & $ 0 $ & $
      (\tau_i\cdot\tau_j)(\vs_i\times\vs_j)\cdot{\bm X}^{(3)}_{ij,-}$
\\
$4 $ & $ -\frac{g_\rho}{2 m_N} h_\rho^1 $ & $ f_\rho(r) $ & $
      \frac{\mu^2}{\Lambda^3_\chi}(C^{\not{\pi}}_2+C^{\not{\pi}}_4)
      $ & $ f_\mu^{\not{\pi}}(r) $ & $
      \frac{\Lambda^2}{\Lambda^3_\chi}(C^{\pi}_2+C^{{\pi}}_4) $ & $
      f_\Lambda(r) $ & $
      (\tau_i+\tau_j)^z(\vs_i-\vs_j)\cdot{\bm X}^{(4)}_{ij,+}$
\\
$5  $ & $  -\frac{g_\rho(1+\kappa_\rho)}{2 m_N}h_\rho^1 $ & $ f_\rho(r)
     $ & $ 0 $ & $ 0 $ & $
     \frac{2\sqrt{2}\pi g_A^3\Lambda^2}{\Lambda_\chi^3}h^1_\pi $ & $
      L^\pi_\Lambda(r) $ & $
      (\tau_i+\tau_j)^z(\vs_i\times\vs_j)\cdot{\bm X}^{(5)}_{ij,-}$
\\
$6 $ & $ -\frac{g_\rho}{2\sqrt{6} m_N}h_\rho^2 $ & $ f_\rho(r) $ & $
  -\frac{2\mu^2}{\Lambda^3_\chi}C_5^{\not{\pi}} $
  & $   f^{\not{\pi}}_\mu(r)$ & $
  -\frac{2\Lambda^2}{\Lambda^3_\chi}C_5^{{\pi}} $
  & $ f_\Lambda(r)$ & $
   {\cal T}_{ij}
   (\vs_i-\vs_j)\cdot{\bm X}^{(6)}_{ij,+}$
\\
$7 $ & $ -\frac{g_\rho(1+\kappa_\rho)}{2\sqrt{6} m_N}h_\rho^2 $ & $ f_\rho(r) $ & $
   0 $ & $ 0 $ & $
    0  $ & $ 0 $ & $
   {\cal T}_{ij}(\vs_i\times\vs_j)\cdot{\bm X}^{(7)}_{ij,-}$
\\
$8 $ & $ -\frac{g_\omega}{m_N}h_\omega^0 $ & $ f_\omega(r) $ & $
   \frac{2\mu^2}{\Lambda^3_\chi} C_1^{\not{\pi}} $ & $ f_\mu^{\not{\pi}}(r) $ & $
   \frac{2\Lambda^2}{\Lambda^3_\chi} C_1^{{\pi}} $ & $ f_\Lambda(r) $ & $
      (\vs_i-\vs_j)\cdot{\bm X}^{(8)}_{ij,+}$
\\
$9  $ & $  -\frac{g_\omega(1+\kappa_\omega)}{m_N} h_\omega^0 $ & $ f_\omega(r) $ & $
   \frac{2\mu^2}{\Lambda_\chi^3}\tilde{C}^{\not{\pi}}_1 $ & $ f_\mu^{\not{\pi}}(r) $ & $
   \frac{2\Lambda^2}{\Lambda_\chi^3}\tilde{C}^{{\pi}}_1 $ & $ f_\Lambda(r) $ & $
   (\vs_i\times\vs_j)\cdot{\bm X}^{(9)}_{ij,-}$
\\
$10 $ & $ -\frac{g_\omega}{2 m_N} h_\omega^1 $ & $ f_\omega(r) $ & $
     0 $ & $ 0 $ & $
      0  $ & $ 0  $ & $
     (\tau_i+\tau_j)^z(\vs_i-\vs_j)\cdot{\bm X}^{(10)}_{ij,+}$
\\
$11 $ & $ -\frac{g_\omega(1+\kappa_\omega)}{2m_N} h^1_\omega $ & $ f_\omega(r) $ & $
    0 $ & $  0 $ & $
     0  $ & $  0  $ & $
    (\tau_i+\tau_j)^z(\vs_i\times\vs_j)\cdot{\bm X}^{(11)}_{ij,-}$
\\
$12 $ & $ -\frac{g_\omega h_\omega^1-g_\rho h_\rho^1}{2m_N} $ & $ f_\rho(r) $ & $
   0 $ & $ 0 $ & $
    0 $ & $ 0   $ & $
   (\tau_i-\tau_j)^z(\vs_i+\vs_j)\cdot{\bm X}^{(12)}_{ij,+}$
\\
$13 $ & $ -\frac{g_\rho}{2m_N} h^{'1}_\rho $ & $ f_\rho(r) $ & $
      0 $ & $ 0 $ & $
       -\frac{\sqrt{2}\pi g_A\Lambda^2}{\Lambda_\chi^3} h_\pi^1 $ & $ L^\pi_\Lambda(r) $ & $
     (\tau_i\times\tau_j)^z(\vs_i+\vs_j)\cdot{\bm X}^{(13)}_{ij,-}$
\\
$14$ & 0  & 0 &  0 & 0 &
 $\frac{2\Lambda^2}{\Lambda^3_\chi}   C^{{\pi}}_6$ &
 $f_\Lambda(r)$ &
 $(\tau_i\times\tau_j)^z(\vs_i+\vs_j)\cdot{\bm X}^{(14)}_{ij,-}$
\\
$15$   &  0  &  0   & 0 &  0   &
   $\frac{\sqrt{2}\pi g^3_A\Lambda^2}{\Lambda_\chi^3} h_\pi^1 $ &
   $\tilde{L}^\pi_\Lambda(r)$ &
   $(\tau_i\times\tau_j)^z(\vs_i+\vs_j)\cdot{\bm X}^{(15)}_{ij,-}$
\end{tabular}
\end{ruledtabular}
\end{table}

The  operators $O^{(n)}_{ij}$ in the last column are represented as products of isospin, spin,  and vector operators ${\bm X}^{(n)}_{ij,\pm}$, which are
defined as
\bea
{\bm X}^{(n)}_{ij,+} & \equiv&[\vp_{ij},f_n(r_{ij})]_{+},\no
{\bm X}^{(n)}_{ij,-}&\equiv&i [\vp_{ij},f_n(r_{ij})]_{-},
\eea
where $\vp_{ij}\equiv\frac{(\vp_i-\vp_j)}{2}$.

One can see that all weak potentials have the same structure, being represented by the 15  basic operators which are  allowed by  symmetry.
Thus, the difference between the weak potentials
is due merely to the choice of coupling constants assigned to each operator  and the scalar functions which describe the radial behavior of the term with the particular operator.
This leads to a principle difference between the DDH and EFT-type of weak potentials, because the scalar functions for the DDH potential are well constrained   by a specific  meson exchange forces, while the choice for the scalar functions in EFT approach is rather arbitrary.
Therefore, EFT-potentials have more degrees of freedom, which leads to the possibility to parameterize a larger set of independent observables. However, one cannot predict the values of parity violating effects using EFT-potentials unless all LECs are determined.

For the case of the DDH potential, radial functions $f_{x}(r)$, $x=\pi,\rho$, and $\omega$ are usually written as normal Yukawa functions or
modified Yukawa functions with corresponding cutoff terms
\bea
f_{x}(r)=\frac{1}{4\pi r}\left\{e^{-m_x r}-e^{-\Lambda_x r}\left[
                          1+\frac{\Lambda_x r}{2}\left(1-\frac{m_x^2}{\Lambda_x^2}\right)
                          \right]\right\},
\eea
where, $m_x$ is a $x$-meson mass, and $\Lambda_x$ is a corresponding cutoff parameter.
We adopt two sets of the scalar functions, with and without cutoff terms, as described in Table \ref{tbl:pv:DdhCutValues}, and call them DDH-I and DDH-II.

\begin{table}%[H] add [H] placement to break table across pages
\caption{\label{tbl:pv:DdhCutValues}
The cutoff parameters  for the DDH parity violating potentials in GeV units\cite{Schiavilla:2004wn}.
For the masses of the mesons we use
$m_\pi=0.138$ GeV, $m_\rho=0.771$ GeV, and $m_\omega=0.783$ GeV.
}
\begin{ruledtabular}
\begin{tabular}{cccccccc}
 & $\Lambda_\pi$ & $\Lambda_\rho$ & $\Lambda_\omega$ \\
DDH-I & $1.72$ & $1.31$ & $1.50$ \\
DDH-II & $\infty$ & $\infty$ & $\infty$ \\
\end{tabular}
\end{ruledtabular}
\end{table}

In the EFT, the results of the calculations of low energy observables should be independent
of the specific form of the scalar functions $f_\mu(r)$   in
the pionless EFT  ($\not{\pi}$EFT) potentials
and of the form of the scalar functions used for the contact terms in pionful EFT ($\pi$EFT), provided
these functions are well localized (close to the delta function) and, at the same time, are smooth enough
to be used in numerical calculations.
This is because the dependencies on the  mass scale ($\mu$)   and  on the particular choice of the form of these functions must be absorbed by the renormalization of
the low energy constants.
Then, for our calculations in pionless EFT, we use two sets of the scalar functions,
 which we call $\not{\pi}$EFT-I and $\not{\pi}$EFT-II,
respectively \footnote{Note that these terms are different
from ones used in \cite{Song:2010sz}.}:
\bea
f_\mu(r)&=&\frac{1}{4\pi r}e^{-\mu r},\mbox{ for $\not{\pi}$EFT-I},\no
f_\Lambda(r)&=&\frac{1}{\Lambda^2}\delta_{\Lambda}(r)
     =\frac{1}{\Lambda^2}\int \frac{d^3 k}{(2\pi)^3}e^{-i\vk\cdot\vr}
      e^{-\frac{k^2}{\Lambda^2}},\mbox{for $\not{\pi}$EFT-II}
\eea
with mass scale parameters $\mu$ and $\Lambda$
which provide a cutoff scale of the
theory. For example,
the natural scale  of the cutoff parameters in pionless theory
is $(\mu,\Lambda)\simeq m_\pi$.

The pionful EFT model ($\pi$EFT) has  explicit
 long range interaction terms
resulting from one pion exchange ($V_{-1,LR}$) and higher order long range
corrections ($V_{1,LR}$). Also, it has
middle range interactions due to the two pion exchange ($V_{1,MR}$),
as well as a short range interactions ($V_{1,SR}$)
due to nucleon contact terms.
The radial part of the leading term of
the long range one pion exchange, $V_{-1,LR}$,
is described by the modified Yukawa function $f_\pi(r)$.
The short range interaction function $V_{1,SR}$ in pionful theory
has the same structure as  pionless EFT.
However,  in spite of the structural similarity,
the origins for these functions are different, therefore, as a consequence,   their numerical
values can be different.
The only term in pionful EFT which has a different operator structure as compared to DDH or pionless EFT potentials, is a higher order long range correction term  $V^{PV}_{1,LR}$.
However, we can ignore these higher order corrections
from  long range interactions, because they are suppressed and
 can  be absorbed by renormalization of
low energy constants \cite{Liu:2006dm}.
Therefore, pionful EFT does not introduce a
new operator structure as long as
we neglect $V^{PV}_{1,LR}$ term \cite{Liu:2006dm, Hyun:2006mp}.
The middle range interactions $V_{1,MR}$, or two pion exchange,
can be described by functions $L(q)$ and $H(q)$ in momentum space
\bea
L(q)\equiv \frac{\sqrt{4 m_\pi^2+\vq^2}}{|\vq|}\ln\left(
                         \frac{\sqrt{4 m_\pi^2+\vq^2}+|\vq|}{2m_\pi}\right),
\quad
H(q)\equiv \frac{4 m_\pi^2}{4 m_\pi^2+\vq^2}L(q),
\eea
where, $q^\mu=(q^0,\vq)=p_1^\mu-p^{'\mu}_1=p^{'\mu}_2-p_2^\mu$.
To transform these scalar functions into a configuration space representation by Fourier transform, we use the regulators $S_\Lambda(q)$.
For the sake of simplicity, we use only one cutoff parameter for
all regulators.
%, both for middle range and for short range interactions.
Therefore, one can write
\bea
\{L_\Lambda(r),H_\Lambda(r),f_\Lambda(r),f^\pi_\Lambda(r)\}
=\frac{1}{\Lambda^2}
\int \frac{d^3 q}{(2\pi)^3} e^{-i\vq\cdot\vr}
S_\Lambda(q)\{L(q),H(q),1,\frac{\Lambda^2}{\vq^2+m_\pi^2}\},
\eea
where $L_\Lambda(r)$ and $H_\Lambda(r)$ correspond to two-pion exchange
loop contributions, $f_\Lambda(r)$ and $f^\pi_\Lambda(r)$ describe short range contact
terms and long range one-pion exchange
contributions, correspondingly. It should be noted that we introduce the cutoff function even for the case
of long range one-pion exchange potential to regularize a short range
part of one-pion exchange.
Among the possible choices, we use two types of
regulators, which are called $\pi$EFT-I
and $\pi$EFT-II\footnote{Note that the convention is different  from the one used in
\cite{Song:2010sz}.}:
\bea
S_\Lambda^{exp}(\vq)=e^{-\frac{\vq^2}{\Lambda^2}}\mbox{ for $\pi$EFT-I},
\quad
S_\Lambda^{dipole}(\vq)=\frac{(\Lambda^2-4m_\pi^2)^2}{(\Lambda^2+\vq^2)^2}
\mbox{ for $\pi$EFT-II}.
\eea
One can see that the function $f_\Lambda(r)$ in $\pi$EFT-I looks similar to the function for $\not{\pi}$EFT-II case; however,
it leads to a different regularization since the
typical value of the cutoff parameter
for $\pi$EFT theory exceeds the pion mass scale and
should be at least about of the $\rho$ meson mass scale,
while for pionless  case it is close to the pion mass.
Therefore, LECs for  the same operators in pionless  and pionful EFT potentials
can be very different.

\subsection{Three nucleon wave functions\label{sec:FE}}

Nuclear wave functions of initial (neutron-deuteron scattering) and final (bound triton)
states of the neutron-deuteron radiative capture process are
obtained in the context of non-relativistic quantum  three particle problem.
We consider neutrons and protons as the isospin degenerate states of the same particle nucleon, whose
mass is fixed to $\hbar ^{2}/m=41.471$ MeV$\cdot$fm. The three-particle problem is formulated
by means of Faddeev equations in configuration space~\cite{Faddeev:1960su}. Using  isospin formalism,
three Faddeev equations become formally identical, which for pairwise interactions reads
\begin{equation}
\left( E-H_{0}-V_{ij}\right) \psi _{k}
=V_{ij}(\psi _{i}+\psi _{j}),
\label{EQ_FE}
\end{equation}
where $(ijk)$ are particle indexes, $H_{0}$ is kinetic energy operator,
$V_{ij}$ is a two body force between particles $i$, and $j$,
$\psi _{k}=\psi_{ij,k}$ is so called Faddeev component.
In the last equation, the potential formally contains both strong
interaction, parity conserving, part ($V^{PC}_{ij}$) and weak interaction, parity
violating, part ($V^{PV}$), i.e.: $V_{ij}=V^{PC}_{ij}+V^{PV}_{ij}$.
Due to the presence of
parity violating potential, the system's wave function does not have a definite parity and contains both
 positive and negative parity components.
As a consequence, the Faddeev components of the total wave function can be split in to the sum of positive and negative parity parts:
\begin{equation}
\psi _{k}=\psi^{+}_{k}+\psi^{-}_{k}
\end{equation}
At  low neutron energies, the dominant components of   both initial  and final state nuclear
wave functions have  positive parity.
Parity violating interaction is weak ($V^{PV}_{ij}<<V^{PC}_{ij}$), then by neglecting second order weak potential
terms  one obtains a system of two differential equations:
\begin{eqnarray}
\left( E-H_{0}-V^{PC}_{ij}\right) \psi^{+} _{k}
&=&V^{PC}_{ij}(\psi^{+} _{i}+\psi^{+} _{j}) \label{EQ_FE_1}, \\
\left( E-H_{0}-V^{PC}_{ij}\right) \psi^{-} _{k}
&=&V^{PC}_{ij}(\psi^{-} _{i}+\psi^{-} _{j})+V^{PV}_{ij}(\psi^{+} _{i}+\psi^{+} _{j}+\psi^{+} _{k}) \label{EQ_FE_2}
\end{eqnarray}
One can see that the
first equation (\ref{EQ_FE_1}) defines only the positive parity part of the wave function.
This equation contains only a strong
nuclear potential and corresponds to the standard three nucleon problem: s-wave  neutron-deuteron scattering,  or an bound state of the triton.
The solution of the second differential equation
(\ref{EQ_FE_2}), which contains inhomogeneous term $V^{PV}_{ij}(\psi^{+} _{i}+\psi^{+} _{j}+\psi^{+} _{k})$, gives us negative parity components of wave functions.

To solve these equations numerically, we use our standard procedure, described in  detail in~\cite{These_Rimas_03}.
Using a set of Jacobi coordinates, defined by
$\vx_{k}=(\vr_{j}-\vr_{i})\smallskip $
and
$\vy_{k}=
\frac{2}{\sqrt{3}}(\vr_{k}-\frac{\vr_{i}+\vr_{j}}{2})$, we expand each Faddeev component of the wave function
 in bipolar harmonic basis:
\begin{equation}
\psi _{k}^{\pm}=\sum\limits_{\alpha }\frac{F^{\pm}_{\alpha }(x_{k},y_{k})}{x_{k}y_{k}}%
\left\vert \left( l_{x}\left( s_{i}s_{j}\right) _{s_{x}}\right)
_{j_{x}}\left( l_{y}s_{k}\right) _{j_{y}}\right\rangle _{JM}\otimes
\left\vert \left( t_{i}t_{j}\right) _{t_{x}}t_{k}\right\rangle _{TT_{z}},
\label{EQ_FA_exp}
\end{equation}%
where index $\alpha $ represents all allowed combinations of the
quantum numbers presented in the brackets, $l_{x}$ and $l_{y}$ are the
partial angular momenta associated with respective Jacobi coordinates, $%
s_{i} $ and $t_{i}$ are  spins and isospins of the individual particles. Functions
$F_{\alpha }(x_{k},y_{k})$ are called partial Faddeev amplitudes.
It should be noted that the total angular momentum $J$, as well as its
projection $M$, are conserved.  Isospin breaking is taken fully into account by
considering both $T=1/2$ and $T=3/2$  channels of the total isospin.

Equations (\ref{EQ_FE_1}) and (\ref{EQ_FE_2}) must be supplemented with the appropriate boundary conditions for
Faddeev partial amplitudes $F_{\alpha }^{\pm}$.
First of all, partial Faddeev amplitudes are regular at the origin:
\begin{equation}
F_{\alpha }^{\pm}(0,y_{k})=F^{\pm}_{\alpha }(x_{k},0)=0.  \label{BC_xyz_0}
\end{equation}%
For a bound state problem, system's wave function also vanish exponentially as either $x_{k}$
or $y_{k}$ becomes large. This condition is imposed by setting Faddeev amplitudes to vanish
at the borders $(x_{max},y_{max})$ of  a chosen grid , i.e.:
\begin{equation}
F_{\alpha }^{\pm}(x_{k},y_{max})=0,\quad
F^{\pm}_{\alpha }(x_{max},y_{k})=0.  \label{BC_xyz_0}
\end{equation}%
For neutron-deuteron scattering
with energies below the break-up threshold, partial Faddeev amplitudes
also vanish  for $\mathbf{x}_{k}\rightarrow\infty$, thus the last equality in (\ref{BC_xyz_0})
also applies for the scattering.

At $\mathbf{y}_{k}\rightarrow\infty $, all the Faddeev amplitudes vanish except for those
consistent with the open channel, describing neutron-deuteron relative motion. For the case of thermal neutrons,
we keep only the relative s-wave amplitudes in the asymptote. This behavior is imposed by:
\bea
F_{\alpha}^{(\pm)}(x,y_{max})= f^{(\pm)}_{l_x,j_x,s_x,t_x}(x)(y_{max}-\frac{2}{\sqrt{3}}a_J)\delta_{l_y,0}\delta_{j_y,1/2}\delta_{j_x,1}.
\eea
Here, $f^{(\pm)}_{l_x,j_x}(x)$  are reduced deuteron wave function components with respective parity $(\pm)$,
orbital momenta $l_x$, total angular momentum $j_x$, total spin $s_x$ and total isospin $t_x$.
 The corresponding deuteron wave function is calculated  before three-nucleon scattering problem is undertaken. Neutron-deuteron scattering lengths
$a_J$ for an angular momenta $J=1/2$ and $J=3/2$ are obtained by solving equation~(\ref{EQ_FE_1}).

The formalism described above can be easily generalized to accommodate three-nucleon forces, as
described in paper~\cite{Lazauskas_2008}.

\subsection{Evaluation of the matrix elements}

 In order to calculate the parity violating $E1$ matrix elements,
 we define real
$\widetilde{\cal E}^{(n)}_J$ matrix elements corresponding
to each operator $O^{(n)}$ as
\bea
\widetilde{\cal E}_J=\sum_n c_n \widetilde{\cal E}^{(n)}_J ,
\eea
where the sum is taken over different parity violating operators with corresponding LECs $c_n$,  defined in the table \ref{tbl:pvpotential}.
At the leading order, the electromagnetic charge operator
does not violate parity.  Therefore, the parity violating E1 amplitude results only from the
small admixture of the parity violating
component of wave functions.
In the convention we use the parity violating wave functions
are purely imaginary both for bound state as well as for zero energy n-d scattering, one has
\bea
\widetilde{\cal E}^{(n)}_J=-\widetilde{\cal E}^{(n)}_{J,(+)}
                     +\widetilde{\cal E}^{(n)}_{J,(-)},
\eea
where $\widetilde{\cal E}_{J,(\pm)}$ are amplitudes for transitions from a
parity conserving scattering wave to a parity violating bound state,
and from a parity violating scattering wave to a parity conserving bound state, respectively.

In the first order of perturbation,
 parity violating $E1$ amplitudes can be presented as a linear combination of
  matrix elements $X^{(m)}$ calculated for each of the
parity violating potential operators $O^{(m)}_{ij}$. Then,
all PV observables $a^\gamma_n$, $P^\gamma$, $A_d^\gamma$ can be expanded as
\bea
X&=&\sum \left(\frac{c_m}{\mu_N}\right) X^{(m)},
\eea
(where $X$ stands for $a^\gamma_n$, $P^\gamma$, or $A_d^\gamma$,
and $\mu_N$ is introduced because of a dimension of coefficients $c_m$)
 in terms of corresponding multipole amplitudes $X^{(m)}$, presented by the following expressions:
\bea
a_n^{\gamma,(m)} &=&(-\frac{2}{3}\sqrt{4\pi})
 \frac{ \left[
   \sqrt{2}(\widetilde{\cal E}^{(m)}_{\frac{3}{2}} \widetilde{\cal M}_{\frac{1}{2}}
        +\widetilde{\cal E}^{(m)}_{\frac{1}{2}} \widetilde{\cal M}_{\frac{3}{2}})
   +\frac{5}{2}(\widetilde{\cal E}^{(m)}_{\frac{3}{2}} \widetilde{\cal M}_{\frac{3}{2}})
   -(\widetilde{\cal E}^{(m)}_{\frac{1}{2}}\widetilde{\cal M}_{\frac{1}{2}})
   \right]}
        {|\widetilde{\cal M}_{\frac{1}{2}}|^2+|\widetilde{\cal M}_{\frac{3}{2}}|^2}
\eea
\bea
P^{\gamma,(m)}
&=&(-2\sqrt{4\pi}) \frac{
\left[\widetilde{\cal E}^{(m)}_{\frac{1}{2}}\widetilde{\cal M}_{\frac{1}{2}}
  +\widetilde{\cal E}^{(m)}_{\frac{3}{2}}\widetilde{\cal M}_{\frac{3}{2}}\right]
}{|\widetilde{\cal M}_{\frac{1}{2}}|^2
                   +|\widetilde{\cal M}_{\frac{3}{2}}|^2
                    }
\eea
\bea
A_d^{\gamma,(m)}
&=&(\frac{1}{2}\sqrt{4\pi})
            \frac{\left[-5 \widetilde{\cal E}^{(m)}_{\frac{3}{2}} \widetilde{\cal M}_{\frac{3}{2}}
                     -4 \widetilde{\cal E}^{(m)}_{\frac{1}{2}} \widetilde{\cal M}_{\frac{1}{2}}
                    +\sqrt{2}\widetilde{\cal E}^{(m)}_{\frac{3}{2}} \widetilde{\cal M}_{\frac{1}{2}}
                    +\sqrt{2}\widetilde{\cal E}^{(m)}_{\frac{1}{2}} \widetilde{\cal M}_{\frac{3}{2}}
\right]}{(|\widetilde{\cal M}_{\frac{1}{2}}|^2+|\widetilde{\cal M}_{\frac{3}{2}}|^2)}.
\eea

It should be noted that for EFT potentials, each parity violating coefficient $c_n$ has an explicit cutoff or scale
dependence multiplier $\frac{1}{\mu^2}(\mbox{ or }\frac{1}{\Lambda^2})$. Therefore, we present all results in normalized forms,  as $\mu^2(\mbox{or } \Lambda^2)\times
 \widetilde{\cal E}^{(m)}(\mbox{or } X^{(n)})$,
 to remove this artificial scale dependence.

We calculate the parity violating E1 amplitude
using 1-body charge operator
\bea
E1_J=\la J_B || \frac{q}{\sqrt{6\pi}}\sum_i Q_i r_i || J\ra
=(-i)\sum_n \frac{\omega}{\sqrt{6\pi}} c_n \widetilde{\cal E}^{(n)}_J,
\eea
where, $Q_i$ and $r_i$ are i-th nucleons charge and position
in the center of mass
system, such that
\bea
\sum_{i=1}^3 Q_i \vr_i
 &=&\frac{1}{2}\left(\frac{1}{2}\vx_3(\tau_2-\tau_1)^z
   +\frac{1}{\sqrt{3}}\vy_3(\tau_3-\frac{\tau_1+\tau_2}{2})^z\right).
\eea
Then, using the wave function expansion
\bea
|\psi_i\ra=\sum_{\alpha} \frac{F_{\alpha,i}(x,y)}{xy}|\alpha\ra ,
\eea
one obtains
\bea
E1&=&\sqrt{\frac{1}{6\pi}}\omega
   (\sqrt{\frac{3}{4}})^3 \sum_{\alpha,\beta}
   \int dx x^2 \int dy y^2 \left(
   \frac{F^*_{\beta,f}(x,y)}{xy}\frac{1}{4}x\frac{F_{\alpha,i}(x,y)}{xy}
   \la \beta||\hat{x}||\alpha\ra\la \beta|(\tau_2-\tau_1)^z|\alpha\ra
   \right.\no & &\left.
   +\frac{F^*_{\beta,f}(x,y)}{xy}\frac{1}{2\sqrt{3}}y
    \frac{F_{\alpha,i}(x,y)}{xy}
    \la \beta||\hat{y}||\alpha\ra
    \la\beta|\left(\tau_3-\frac{\tau_1+\tau_2}{2}\right)^z|\alpha\ra
   \right),
\eea
where $(\sqrt{\frac{3}{4}})^3$ comes from the normalization of $y$.
For these amplitudes, the integration over radial function is done numerically but
angular parts of matrix elements are calculated analytically.

\section{Results and Discussions}
The results of our calculations are presented separately for three choices of weak potentials: for the DDH potential, for the pionless and for the pionful potentials derived in the EFT approach.

\subsection{The DDH potential results}

The results obtained with the DDH potential are in a reasonably good agreement with the previous calculations \cite{Moskalev:1969,Hadjimichael:1974,McKellar:1974yr,Desplanques:1986cq},
considering the difference in wave functions,
  and  give us the opportunity to estimate the values of all PV effects in terms of PV meson-nucleon coupling constants $h$ as
\begin{eqnarray}\label{param}
a_n &=&  0.42  h^{1}_{\pi}
 -0.17  h^{0}_{\rho}
 +0.085 h^{1}_{\rho}
 +0.008 h^{2}_{\rho}
 -0.238	h^{0}_{\omega}
 +0.086 h^{1}_{\omega}
 -0.010 h^{\prime 1}_{\rho}
=4.11\times 10^{-7}        \\
P_\gamma&=& -1.05  h^{1}_{\pi}
 +0.19  h^{0}_{\rho}
 -0.096 h^{1}_{\rho}
 -0.018 h^{2}_{\rho}
 +0.28  h^{0}_{\omega}
 -0.046 h^{1}_{\omega}
 +0.023 h^{\prime 1}_{\rho}
= -7.31\times 10^{-7}  \\
A_d^\gamma&=& -1.51       h^{1}_{\pi}
 +0.17       h^{0}_{\rho}
 -0.083      h^{1}_{\rho}
 -0.024      h^{2}_{\rho}
 +0.024      h^{0}_{\omega}
 +0.013      h^{1}_{\omega}
 +0.032      h^{\prime 1}_{\rho}
=-9.05\times 10^{-7}.
\end{eqnarray}

The coefficients in these expressions are obtained using strong  AV18+UIX  and weak DDH-II potentials, while the final values of PV observables are given for the ``best'' values of DDH coupling constants.
The contributions of different PV operators to transition amplitudes
$\widetilde{\cal E}_{J,(P)}$, where
   $(P)$ indicates the parity of the  scattering waves,  are shown in table \ref{tbl:matv:DDH2:Av18u9}. One can see that unlike  the n-d elastic scattering case, there is no dominance of $J=\frac{3}{2}$ channel and, as a consequence,  all operators contribute almost equally to the capture process.

\begin{table}[H]%[H] add [H] placement to break table across pages
\caption{\label{tbl:matv:DDH2:Av18u9}
Parity violating amplitudes
$\widetilde{\cal E}_{J,(P)}$
  in fm$^{\frac{3}{2}}$ units, where $(P)$ stands for the parity of the
  scattering wave, calculated with
 AV18+UIX strong  and  DDH-II weak
potentials.
}
\begin{ruledtabular}
\begin{tabular}{rrrrr}
 n    & $\widetilde{\cal E}_{\frac{1}{2},(+)}$
      & $\widetilde{\cal E}_{\frac{1}{2},(-)}$
      & $\widetilde{\cal E}_{\frac{3}{2},(+)}$
      & $\widetilde{\cal E}_{\frac{3}{2},(-)}$  \\
 \hline
  1 &$-3.37\times 10^{-1}$&$-3.75\times 10^{-2}$&$-1.44\times 10^{-2}$&$-2.97\times 10^{-1}$\\
  2 &$-2.64\times 10^{-3}$&$-1.52\times 10^{-2}$&$-5.37\times 10^{-3}$&$-2.52\times 10^{-2}$\\
  3 &$-9.72\times 10^{-3}$&$ 3.12\times 10^{-2}$&$-1.35\times 10^{-2}$&$ 1.31\times 10^{-2}$\\
  4 &$ 1.03\times 10^{-2}$&$-1.32\times 10^{-2}$&$ 1.47\times 10^{-2}$&$-2.87\times 10^{-3}$\\
  5 &$ 1.26\times 10^{-2}$&$-1.56\times 10^{-2}$&$ 1.75\times 10^{-2}$&$-3.79\times 10^{-3}$\\
  6 &$-2.03\times 10^{-3}$&$-8.85\times 10^{-3}$&$-1.85\times 10^{-3}$&$ 1.51\times 10^{-3}$\\
  7 &$-2.42\times 10^{-3}$&$-9.62\times 10^{-3}$&$-2.45\times 10^{-3}$&$ 1.94\times 10^{-3}$\\
  8 &$-7.37\times 10^{-3}$&$ 2.43\times 10^{-2}$&$-1.08\times 10^{-2}$&$ 9.51\times 10^{-3}$\\
  9 &$-7.10\times 10^{-3}$&$ 1.24\times 10^{-2}$&$-1.05\times 10^{-2}$&$-2.14\times 10^{-3}$\\
 10 &$ 9.79\times 10^{-3}$&$-1.25\times 10^{-2}$&$ 1.39\times 10^{-2}$&$-2.71\times 10^{-3}$\\
 11 &$ 1.20\times 10^{-2}$&$-1.48\times 10^{-2}$&$ 1.67\times 10^{-2}$&$-3.61\times 10^{-3}$\\
 12 &$-2.75\times 10^{-3}$&$ 9.29\times 10^{-3}$&$-4.10\times 10^{-4}$&$-9.10\times 10^{-3}$\\
 13 &$-3.05\times 10^{-3}$&$ 1.84\times 10^{-2}$&$-1.96\times 10^{-3}$&$-1.53\times 10^{-2}$\\
\end{tabular}
\end{ruledtabular}
\end{table}

\begin{table}[H] %add [H] placement to break table across pages
\caption{\label{tbl:LEC:DDH}
The DDH PV coupling constants in units of $10^{-7}$ ($h'_\rho$ contribution is neglected).
Strong interactions parameters are
$\frac{g^2_\pi}{4\pi}=13.9$,
$\frac{g^2_\rho}{4\pi}=0.84$,
$\frac{g^2_\omega}{4\pi}=20$,
$\kappa_\rho=3.7$, and
$\kappa_\omega=0$.
}
\begin{ruledtabular}
\begin{tabular}{ccc}
DDH Coupling& DDH `best' & 4-parameter fit\cite{Bowman}
                         %& 3-parameter fit\cite{Bowman}
                         \\
\hline
$h^1_\pi$ &  $+4.56$      &  $-0.456$ %&  $-0.5$
\\
$h_\rho^0$ & $-11.4$      &  $-43.3$  %& $-33$
\\
$h_\rho^2$ & $-9.5$       & $37.1$   %& $41$
\\
$h_\omega^0$&  $-1.9$     & $13.7$  %& $0$
\\
$h_\rho^1$ &   $-0.19$    &  $-0.19$ %& $-0.19$
\\
$h_\omega^1$ & $-1.14$    &  $-1.14$ %& $-1.14$
\\
\end{tabular}
\end{ruledtabular}
\end{table}

\begin{table}[H]%[H] add [H] placement to break table across pages
\caption{\label{tbl:summary:DDH:bestvalues}
Parity violating observables for different  potential models
with the DDH-best parameter values and Bowman's 4-parameter fits
in $10^{-7}$ units.
}
\begin{ruledtabular}
\begin{tabular}{c |ccc |ccc}
           & DDH-best values&       &
           & 4-parameter fits&      &   \\
 models    &${a}_n$ & ${P}_\gamma$ &   ${A}_d$
           &${a}_n$ & ${P}_\gamma$ &   ${A}_d$\\
 \hline
AV18+UIX/DDH-I &$ 3.30$&$-6.38$&$-8.23$&$ 1.97$&$-2.16$&$-1.81$\\
AV18/DDH-II   &$ 4.61$&$-8.30$&$-10.3$&$ 4.60$&$-5.18$&$-4.46$\\
AV18+UIX/DDH-II &$ 4.11$&$-7.30$&$-9.04$&$ 4.14$&$-4.71$&$-4.09$\\
Reid/DDH-II   &$ 4.74$&$-8.45$&$-10.4$&$ 4.70$&$-5.25$&$-4.46$\\
NijmII/DDH-II  &$ 4.71$&$-8.45$&$-10.5$&$ 4.76$&$-5.26$&$-4.41$\\
INOY/DDH-II   &$9.24$&$ -12.9$&$ -13.8$&$17.5$&$-17.9$&$-13.5$\\
\end{tabular}
\end{ruledtabular}
\end{table}

To check the possible model dependence of these results,
we  compare PV observables for the ``best'' DDH values
and for the 4-parameter fit \cite{Bowman} of weak coupling constants
(see table \ref{tbl:LEC:DDH}). For weak potentials, we used both DDH-I and DDH-II radial
functions with strong interactions described by  AV18, AV18+UIX, Reid, NijmII, and INOY models.
The results for these calculations  are summarized in table \ref{tbl:summary:DDH:bestvalues}.
 The difference in the values of PV effects for the ``best'' DDH values
and for the 4-parameter fit proves that the PV effects in the radiative capture
are very sensitive to the particular choice of the values of meson-nucleon  coupling constants.
We observed rather significant model dependence of the individual matrix elements.
This model dependence indicates a possible serious problem in the calculation of PV effects in nuclei, they
require more thorough analysis
of our calculations for EFT-type potentials presented in that follows.

\subsection{Pionless EFT potential results}

 We start to analyze the EFT approach with the PV potentials obtained in pionless EFT by using scalar functions corresponding to two different schemes for cutoff procedure:   $\not{\pi}$EFT-I and $\not{\pi}$EFT-II.
Calculated PV amplitudes for these two weak EFT potentials  for the same AV18+UIX
strong interaction model are summarized in tables \ref{tbl:pionless1:av18u9:138MeV} and \ref{tbl:pionless2:av18u9:138MeV}.  The difference in $\not{\pi}$EFT-I and $\not{\pi}$EFT-II
results is not surprising because they have different forms
of the scalar functions,  the correct comparison of the results should be done
for the products of these amplitudes with corresponding low energy constants. Then, the  renormalization of the LECs  can absorb the differences in the  amplitudes.
Unfortunately, we do not have enough experimental data to obtain these LECs.

\begin{table}[H]%[H] add [H] placement to break table across pages
\caption{\label{tbl:pionless1:av18u9:138MeV}
         Parity violating amplitudes $\widetilde{\cal E}_{J,(P)}$
         for AV18+UIX stong interaction
          and PV $\not{\pi}$EFT-I potential with $\mu=138$ MeV.
}
\begin{ruledtabular}
\begin{tabular}{ccccc}
op&$\tilde{\cal E}_{\frac{1}{2}(+)}$&$\tilde{\cal E}_{\frac{1}{2}(-)}$
  &$\tilde{\cal E}_{\frac{3}{2}(+)}$&$\tilde{\cal E}_{\frac{3}{2}(-)}$\\
  \hline
 1 &$-1.64\times 10^{-1}$&$-1.83\times 10^{-2}$&$-7.04\times 10^{-3}$&$-1.45\times 10^{-1}$\\
 4 &$ 2.68\times 10^{-1}$&$-2.74\times 10^{-1}$&$ 3.99\times 10^{-1}$&$-9.64\times 10^{-2}$\\
 6 &$-6.16\times 10^{-3}$&$-1.96\times 10^{-1}$&$-3.90\times 10^{-2}$&$ 5.30\times 10^{-2}$\\
 8 &$-3.02\times 10^{-1}$&$ 4.07\times 10^{-1}$&$-2.97\times 10^{-1}$&$ 2.18\times 10^{-1}$\\
 9 &$-8.63\times 10^{-2}$&$ 1.74\times 10^{-1}$&$-1.48\times 10^{-1}$&$ 6.34\times 10^{-3}$\\
\end{tabular}
\end{ruledtabular}
\end{table}
\begin{table}[H]%[H] add [H] placement to break table across pages
\caption{\label{tbl:pionless2:av18u9:138MeV}
Parity violating amplitudes $\widetilde{\cal E}_{J,(P)}$
         for AV18+UIX stong interaction
          and PV $\not{\pi}$EFT-II potential with $\mu=138$ MeV.}
\begin{ruledtabular}
\begin{tabular}{ccccc}
op&$\tilde{\cal E}_{\frac{1}{2}(+)}$&$\tilde{\cal E}_{\frac{1}{2}(-)}$
  &$\tilde{\cal E}_{\frac{3}{2}(+)}$&$\tilde{\cal E}_{\frac{3}{2}(-)}$\\
  \hline
  1 &$-5.79\times 10^{-1}$&$-7.66\times 10^{-1}$&$ 4.09\times 10^{-2}$&$-1.36\times 10^{-1}$\\
  4 &$ 4.28\times 10^{-1}$&$ 4.62\times 10^{-2}$&$ 6.81\times 10^{-1}$&$ 1.57\times 10^{-1}$\\
  6 &$ 7.11\times 10^{-2}$&$-1.79\times 10^{-1}$&$-5.55\times 10^{-2}$&$-2.91\times 10^{-2}$\\
  8 &$-5.92\times 10^{-1}$&$ 6.57\times 10^{-3}$&$-5.15\times 10^{-1}$&$ 6.43\times 10^{-2}$\\
  9 &$-1.45\times 10^{-1}$&$ 3.13\times 10^{-1}$&$-3.16\times 10^{-1}$&$ 1.39\times 10^{-1}$\\
\end{tabular}
\end{ruledtabular}
\end{table}

The contributions of different operators from these two weak EFT potentials  with the same AV18+UIX
strong potential to PV effects are shown in tables \ref{tbl:pv:pionless1:av18u9:138MeV} and \ref{tbl:pv:pionless2:av18u9:138MeV}. One can see that  in the pionless EFT, all operators have approximately  the same level of contribution to the PV effects, which is consistent with the
results for the DDH model.

\begin{table}[H]%[H] add [H] placement to break table across pages
\caption{\label{tbl:pv:pionless1:av18u9:138MeV}
Parity violating observables for AV18+UIX strong potential
for $\not{\pi}EFT$-I at $\mu=138$ MeV. The results are in $fm^{-2}$ units.
}
\begin{ruledtabular}
\begin{tabular}{ccrrr}
n& $\frac{c_n}{\mu_N \mu^2}$     &$\mu^2 a_n^{(n)}$  & $\mu^2 P_\gamma^{(n)}$ & $\mu^2 A_d^{(n)}$ \\
\hline
1&$\frac{4 m_N }{\Lambda_\chi^3} C_6^{\not{\pi}}$                   &$ 2.17\times 10^{-2}$&$-5.52\times 10^{-2}$&$-7.93\times 10^{-2}$\\
4&$\frac{2 m_N}{\Lambda_\chi^3} (C_2^{\not{\pi}}+C_4^{\not{\pi}})$  &$-7.94\times 10^{-2}$&$ 6.55\times 10^{-2}$&$ 3.16\times 10^{-2}$\\
6&$-\frac{2}{\Lambda_\chi^3}C_r^{\not{\pi}}$                        &$-2.81\times 10^{-2}$&$ 5.96\times 10^{-2}$&$ 8.01\times 10^{-2}$\\
8&$-\frac{4 m_N }{\Lambda_\chi^3}C_1^{\not{\pi}}$                   &$ 1.04\times 10^{-1}$&$-1.03\times 10^{-1}$&$-7.58\times 10^{-2}$\\
9&$\frac{4 m_N}{\Lambda_\chi^3} \widetilde{C}_1^{\not{\pi}}$        &$ 3.81\times 10^{-2}$&$-4.29\times 10^{-2}$&$-3.67\times 10^{-2}$\\
\end{tabular}
\end{ruledtabular}
\end{table}

\begin{table}[H]%[H] add [H] placement to break table across pages
\caption{\label{tbl:pv:pionless2:av18u9:138MeV}
Parity violating observables for AV18+UIX strong potentials
for $\not{\pi}$EFT-II at $\Lambda=138$ MeV. The results are
in $fm^{-2}$ units.
}
\begin{ruledtabular}
\begin{tabular}{ccrrr}
n& $\frac{c_n}{\mu_N \Lambda^2}$     &$\Lambda^2 a_n^{(n)}$  & $\Lambda^2 P_\gamma^{(n)}$ & $\Lambda^2 A_d^{(n)}$ \\
\hline
1&$\frac{4 m_N }{\Lambda_\chi^3} C_6^{\not{\pi}}$                    &$-2.73\times 10^{-2}$&$ 2.16\times 10^{-2}$&$ 9.19\times 10^{-3}$\\
4&$\frac{2 m_N}{\Lambda_\chi^3} (C_2^{\not{\pi}}+C_4^{\not{\pi}})$   &$-5.56\times 10^{-2}$&$ 2.19\times 10^{-2}$&$-2.32\times 10^{-2}$\\
6&$-\frac{2}{\Lambda_\chi^3}C_r^{\not{\pi}}$                         &$-3.69\times 10^{-2}$&$ 6.53\times 10^{-2}$&$ 8.08\times 10^{-2}$\\
8&$-\frac{4 m_N }{\Lambda_\chi^3}C_1^{\not{\pi}}$                    &$ 8.75\times 10^{-2}$&$-6.76\times 10^{-2}$&$-2.62\times 10^{-2}$\\
9&$\frac{4 m_N}{\Lambda_\chi^3} \widetilde{C}_1^{\not{\pi}}$         &$ 6.71\times 10^{-2}$&$-5.02\times 10^{-2}$&$-1.71\times 10^{-2}$\\
\end{tabular}
\end{ruledtabular}
\end{table}

\subsection{Pionful EFT potential results}

The PV transition amplitudes calculated for  strong  AV18+UIX potential
and PV pionful EFT potential with cutoff parameter
$\Lambda=600$MeV
are presented in table \ref{tbl:pionful1:av18u9:600MeV}.
The results for PV observables are provided in tables \ref{tbl:pv:pionful1:av18u9:600MeV} and
\ref{tbl:pv:pionful2:av18u9:600MeV}.  These tables reveal strong dependence
on the choice of the scalar functions which, as was mentioned in the previous section, are expected to be absorbed by  corresponding LECs.  (For the comparison with pionless  case, one shall take into account additional $\Lambda^2 / \Lambda_\chi^2$ multipliers  in the coefficients of  leading one-pion exchange operators which appear due to loop diagrams contributions in pionful EFT.)

\begin{table}[H]
\caption{ E1 amplitudes calculated for AV18+UIX and $\pi$EFT-I
  at $\Lambda=600$ MeV
 in fm$^{\frac{3}{2}}$ unit. (The $\Lambda^2$ multiplier is not included.)
 }
\label{tbl:pionful1:av18u9:600MeV}
\begin{ruledtabular}
\begin{tabular}{crrrr}
operator &$\tilde{\cal E}_{\frac{1}{2}(+)}$
         &$\tilde{\cal E}_{\frac{1}{2}(-)}$
         &$\tilde{\cal E}_{\frac{3}{2}(+)}$
         &$\tilde{\cal E}_{\frac{3}{2}(-)}$\\
  \hline
 1 &$-3.51\times 10^{-1}$&$-7.40\times 10^{-2}$&$-1.15\times 10^{-2}$&$-2.86\times 10^{-1}$\\
 4 &$ 3.56\times 10^{-2}$&$-3.86\times 10^{-2}$&$ 4.97\times 10^{-2}$&$-7.90\times 10^{-3}$\\
 5 &$ 3.43\times 10^{-2}$&$-4.36\times 10^{-2}$&$ 4.81\times 10^{-2}$&$-9.25\times 10^{-3}$\\
 6 &$-7.37\times 10^{-3}$&$-3.18\times 10^{-2}$&$-6.42\times 10^{-3}$&$ 4.20\times 10^{-3}$\\
 8 &$-2.65\times 10^{-2}$&$ 7.65\times 10^{-2}$&$-3.75\times 10^{-2}$&$ 2.84\times 10^{-2}$\\
 9 &$-2.32\times 10^{-2}$&$ 4.48\times 10^{-2}$&$-3.51\times 10^{-2}$&$-5.06\times 10^{-3}$\\
13 &$-4.32\times 10^{-4}$&$ 6.26\times 10^{-2}$&$-6.46\times 10^{-3}$&$-4.36\times 10^{-2}$\\
14 &$-1.33\times 10^{-2}$&$ 5.33\times 10^{-2}$&$-6.11\times 10^{-3}$&$-4.66\times 10^{-2}$\\
15 &$ 1.27\times 10^{-2}$&$ 1.28\times 10^{-1}$&$-2.19\times 10^{-2}$&$-9.03\times 10^{-2}$\\
\end{tabular}
\end{ruledtabular}
\end{table}

\begin{table}[H]
\caption{PV observables for PV $\pi$EFT-I potential
and AV18+UIX strong potential at $\Lambda=600$ MeV.
}
\label{tbl:pv:pionful1:av18u9:600MeV}
\begin{ruledtabular}
\begin{tabular}{crrr}
operator & $a_n^{\gamma(n)}$ & $P_\gamma^{(n)}$ & $A_d^{\gamma(n)}$\\
\hline
  1 &$ 4.12\times 10^{-2}$&$-1.06\times 10^{-1}$&$-1.53\times 10^{-1}$\\
  4 &$-1.08\times 10^{-2}$&$ 1.03\times 10^{-2}$&$ 7.00\times 10^{-3}$\\
  5 &$-1.14\times 10^{-2}$&$ 1.13\times 10^{-2}$&$ 8.12\times 10^{-3}$\\
  6 &$-3.62\times 10^{-3}$&$ 7.51\times 10^{-3}$&$ 1.00\times 10^{-2}$\\
  8 &$ 1.51\times 10^{-2}$&$-1.63\times 10^{-2}$&$-1.33\times 10^{-2}$\\
  9 &$ 1.00\times 10^{-2}$&$-1.26\times 10^{-2}$&$-1.23\times 10^{-2}$\\
 13 &$ 9.34\times 10^{-3}$&$-2.07\times 10^{-2}$&$-2.83\times 10^{-2}$\\
 14 &$ 9.87\times 10^{-3}$&$-2.20\times 10^{-2}$&$-3.02\times 10^{-2}$\\
 15 &$ 1.70\times 10^{-2}$&$-3.79\times 10^{-2}$&$-5.18\times 10^{-2}$\\
\end{tabular}
\end{ruledtabular}
\end{table}
\begin{table}[H]
\caption{PV observables for PV $\pi$EFT-II potential
and AV18+UIX strong potential at $\Lambda=600$ MeV.}
\label{tbl:pv:pionful2:av18u9:600MeV}
\begin{ruledtabular}
\begin{tabular}{crrr}
operator & $a_n^{\gamma(n)}$ & $P_\gamma^{(n)}$ & $A_d^{\gamma(n)}$\\
\hline
   1 &$ 2.10\times 10^{-2}$&$-5.62\times 10^{-2}$&$-8.20\times 10^{-2}$\\
   4 &$-6.89\times 10^{-2}$&$ 6.53\times 10^{-2}$&$ 4.34\times 10^{-2}$\\
   5 &$-6.44\times 10^{-2}$&$ 6.32\times 10^{-2}$&$ 4.46\times 10^{-2}$\\
   6 &$-2.09\times 10^{-2}$&$ 4.47\times 10^{-2}$&$ 6.03\times 10^{-2}$\\
   8 &$ 9.18\times 10^{-2}$&$-9.83\times 10^{-2}$&$-7.93\times 10^{-2}$\\
   9 &$ 4.97\times 10^{-2}$&$-6.25\times 10^{-2}$&$-6.04\times 10^{-2}$\\
  13 &$ 4.90\times 10^{-2}$&$-1.09\times 10^{-1}$&$-1.49\times 10^{-1}$\\
  14 &$ 2.71\times 10^{-2}$&$-8.36\times 10^{-2}$&$-1.26\times 10^{-1}$\\
  15 &$ 1.10\times 10^{-1}$&$-2.44\times 10^{-1}$&$-3.33\times 10^{-1}$\\
\end{tabular}
\end{ruledtabular}
\end{table}

\subsection{Cutoff and model dependence}

 The presented results reveal the model dependence of the calculated matrix elements,
both on weak as well as on strong interaction. This model dependence has a different
level of importance in calculating PV effects for different approaches.
 For the case of the DDH approach, the model dependence is directly related to the reliability of the calculations of PV effects in nuclei.
In general, the EFT approach shall lead to model independent results; however,
to guarantee the model independence, the intrinsic cutoff dependence must be checked   explicitly.
For the case of a ``hybrid'' EFT approach, which is not completely free from the possible model dependence,
a careful analysis of both cutoff  and model dependence  of matrix elements and physical observables is required.

 In our approach, we used numerically exact wave functions of three-nucleon systems,  however they  depend
 on the choice of the strong Hamiltonian.
 Another possible source of the model dependence is the choice of PV violating potentials, which, for the EFT approach, means a choice of the
 scalar functions used for  the regularization.
It should be noted that in EFT, the model dependence of physical observables is not directly related to the model dependence of the calculated
PV amplitudes because they are affected by the model dependence of the corresponding LECs.
Unfortunately, at the present time these LECs are unknown, which prevents derivation of PV observables.

Since most phenomenological strong potentials  have a similar
long range  behavior, corresponding to one-pion exchange,
the main difference
between  strong potentials is related to the
 middle and the short range contributions.
Thus, rather strong model dependence of PV amplitudes
implies that matrix elements related to
n-d radiative capture process are sensitive to these short range interactions.
This sensitivity to a short range dynamics is a new phenomenon observed  in radiative n-d capture and is in direct contrast with
the case of parity violation in elastic n-d scattering where PV matrix elements are practically insensitive \cite{Song:2010sz}
to the choice of the strong potential.

This is partially related to the fact that in the case of the elastic n-d scattering, the dominant contribution to PV effects comes from the  $J=3/2$ channel, which  is repulsive and  thus less sensitive to the short range details of
the potential.  On contrary,
in the case of n-d radiative capture, almost all channels
contribute  equally to the values of PV effects. In addition to that, for the radiative capture, the mechanism of  pion exchange is not a  dominant one,  and, as a consequence, contributions from heavier meson exchanges (short distance contributions) become important.
Therefore, one can see a number of reasons why  PV three-body
radiative capture processes should be more sensitive to the short distance
dynamics than PV effects in three-body elastic scattering.
It should be noted that even in the two-body case, a circular photon polarization $P^\gamma$ in n-p radiative capture, which is not dominated
by one-pion exchange, shows stronger model dependence \cite{Schiavilla:2004wn} than $a^\gamma_n$, which relies on one-pion exchange contributions.

As it is mentioned above,
strong dependence of PV effects on the choice of
 potentials
could be a serious problem in the case of the DDH meson exchange model,
implying an uncertainty in the theoretical predictions
and a difficulty in comparing results of different calculations.
 On the other hand,
in a regular EFT approach,
dependence on a cutoff and on the choice of a scalar function
must be absorbed-copensated by the renormalization of
the low energy constants.
After the proper renormalization  one must get model independent prediction of the
low energy observables.
This is not exactly true for the hybrid method,
where strong interactions are introduced
by a phenomenological strong potentials.
However, it can be argued that
the short distance details of the system dynamics would not be
very important for the calculations of low energy observables
according to the basic principle
of the effective field theory.
The removal of the model dependence,
related to the difference in short range parts of the wave functions,
can be achieved by the introduction of the cutoff and renormalization of
LECs in hybrid approach.
A study of the behavior of  the calculated matrix elements
as a function of cutoff parameters in hybrid approach
could be used to check the
validity of these arguments.

\begin{figure}[H]
\begin{center}

   \subfigure[$\mu^2\widetilde{\cal E}_{\frac{3}{2}(+)}$ for operator 1]
   {\includegraphics[width=0.4\textwidth]{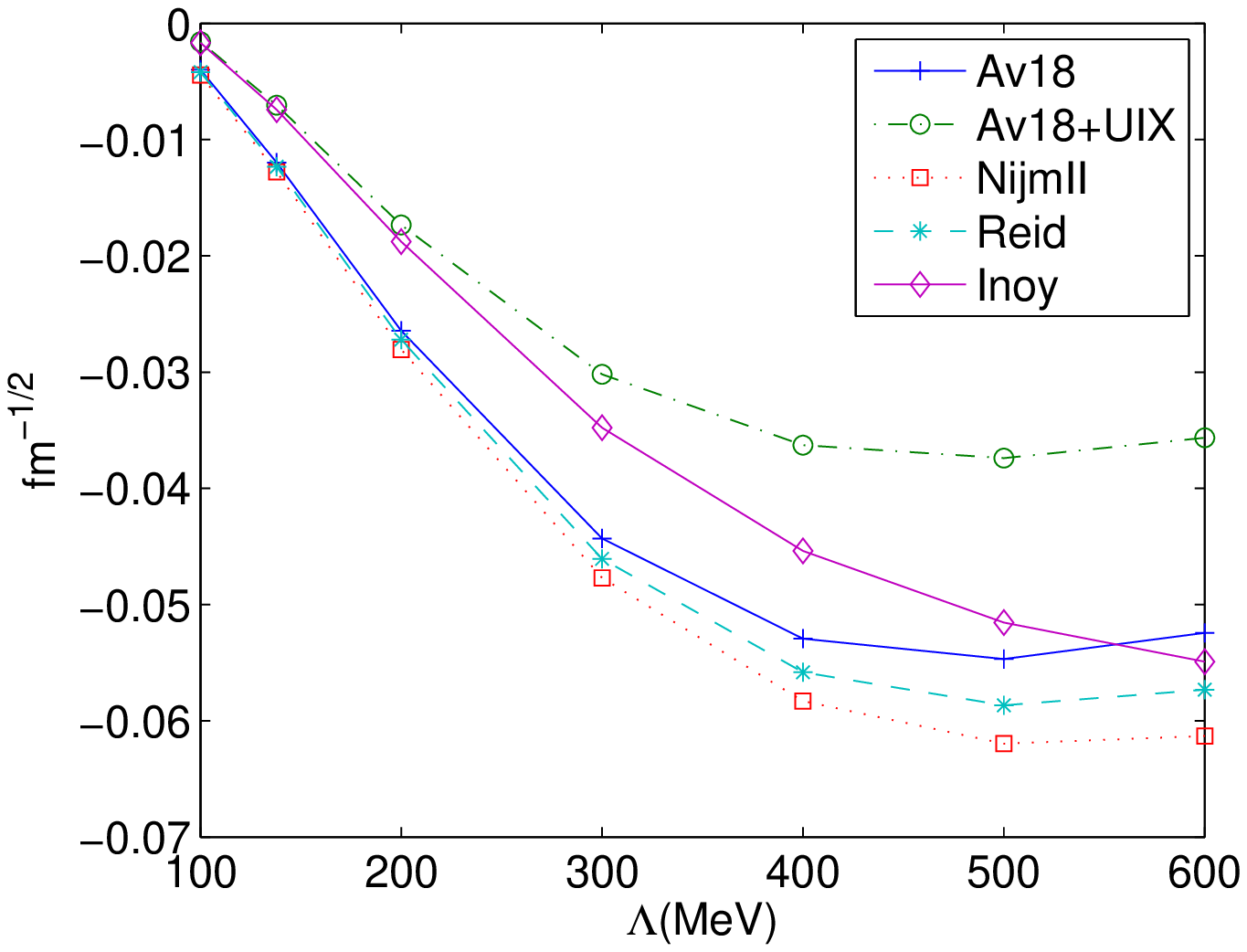}}
   \subfigure[$\mu^2\widetilde{\cal E}_{\frac{3}{2}(+)}$ for operator 9]
   {\includegraphics[width=0.4\textwidth]{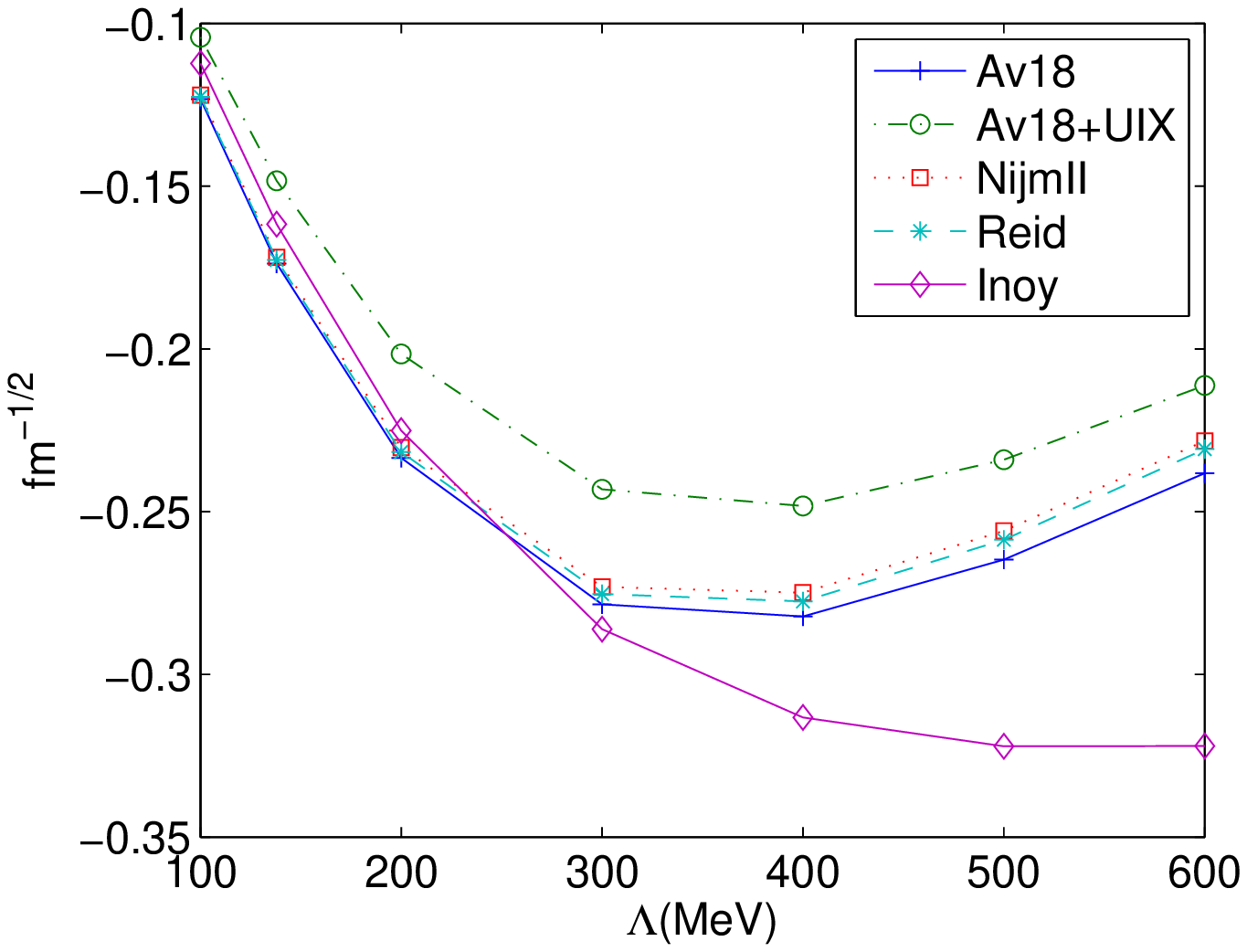}}
    \end{center}
    \caption{\label{fig:pionless1:example}
    Cutoff and strong model dependencies of
    $\mu^2\widetilde{\cal E}$ in $fm^{-\frac{1}{2}}$ for $\not{\pi}$EFT-I calculated with AV18, AV18+UIX, Nijmegen-II, INOY, and Reid
    strong potentials.
    The multiplier $\mu^2$ is used to absorb artificial cutoff dependence of $c_n$
    coefficients.
    }
%\mbox{\epsfxsize=6cm\epsffile{CubicHermite.eps}}
\end{figure}
 As an example, let us consider the $\mu^2\widetilde{\cal E}_{\frac{3}{2}(+)}$ matrix elements as a function of a cutoff mass, which is calculated for  operators 1 and 9 in the $\not{\pi}$EFT-I approach with different strong potentials (see Fig.\ref{fig:pionless1:example}). The choice of these operators is related to their symmetry properties: the operator 1 has quantum numbers corresponding to pion-exchange while the operator 9 - to $\rho$-meson exchange.  It should be noted that since we use the same scalar functions both for the
$\not{\pi}$EFT-I and for the DDH-II schemes of calculations,
we can apply the result of this analysis also to the calculations in the DDH-II scheme.
Once again one observes  rather strong dependence on the choice of a strong potential and
on a cutoff mass parameter.

Analyzing results of Fig.\ref{fig:pionless1:example}
 from the point of view of the DDH approach,
 where the matrix element for the operator 1
at $\mu=m_\pi$ corresponds to the pion-meson exchange and
the matrix element for the operator 9 at $\mu=m_\rho$ corresponds to
the rho-meson exchange,  one can see a large
strong potential model
dependence for heavy meson exchange. This dependence indicates  the
importance of the inclusion of 3-body strong potentials.
Unfortunately, most calculations of PV effects in nuclear physics with the DDH potential
do not include  strong 3-body forces, which could be a possible source for
 the existing discrepancy \cite{Holstein:2006bv} in the analysis of PV effects.

On the other hand, from the point of view of  $\not{\pi}$EFT,
 the reasonable cutoff mass scale cannot exceed the value of the pion mass, where the dependance on strong interaction potential is small.
Since the cutoff in EFT could be considered as a measure of our knowledge of short range physics,
increasing the  cutoff parameter implies stronger
dependence on the short distance details.
Fig.\ref{fig:pionless1:example}
shows that by lowering the cutoff, one can
diminish the strong potential model dependence.
This is because by lowering of the cutoff parameter,  we are effectively switching
to the regime where the theory becomes sensitive only to a long range part of the interaction.
Then, one can expect a smaller
model dependence when the cutoff parameter is low, because all strong potentials
have a similar long range behavior.
Therefore, Fig.\ref{fig:pionless1:example} is consistent with
the basic principle of the EFT and shows that the hybrid method works well.

The  remaining weak  dependence on strong interaction model
at $\mu\simeq m_\pi$ scale could be related both to short and to long range parts of the potentials.
If they are the remainder of the short distance part of wave function,
the difference should be absorbed by LECs. On contrary the difference in the long range
part of the wave function can not be removed by the renormalization of LECs
in the hybrid method. However, as demonstrated in \cite{Song:2008zf},
this  long range part difference is governed by strong
interaction observables and should be easily treated by analyzing the correlation
between matrix elements and effective range parameters.

\begin{figure}[H]
\begin{center}
\subfigure[$\Lambda^2\widetilde{\cal E}_{\frac{1}{2},(+)}$ for op1]
            {\includegraphics[width=0.4\textwidth]{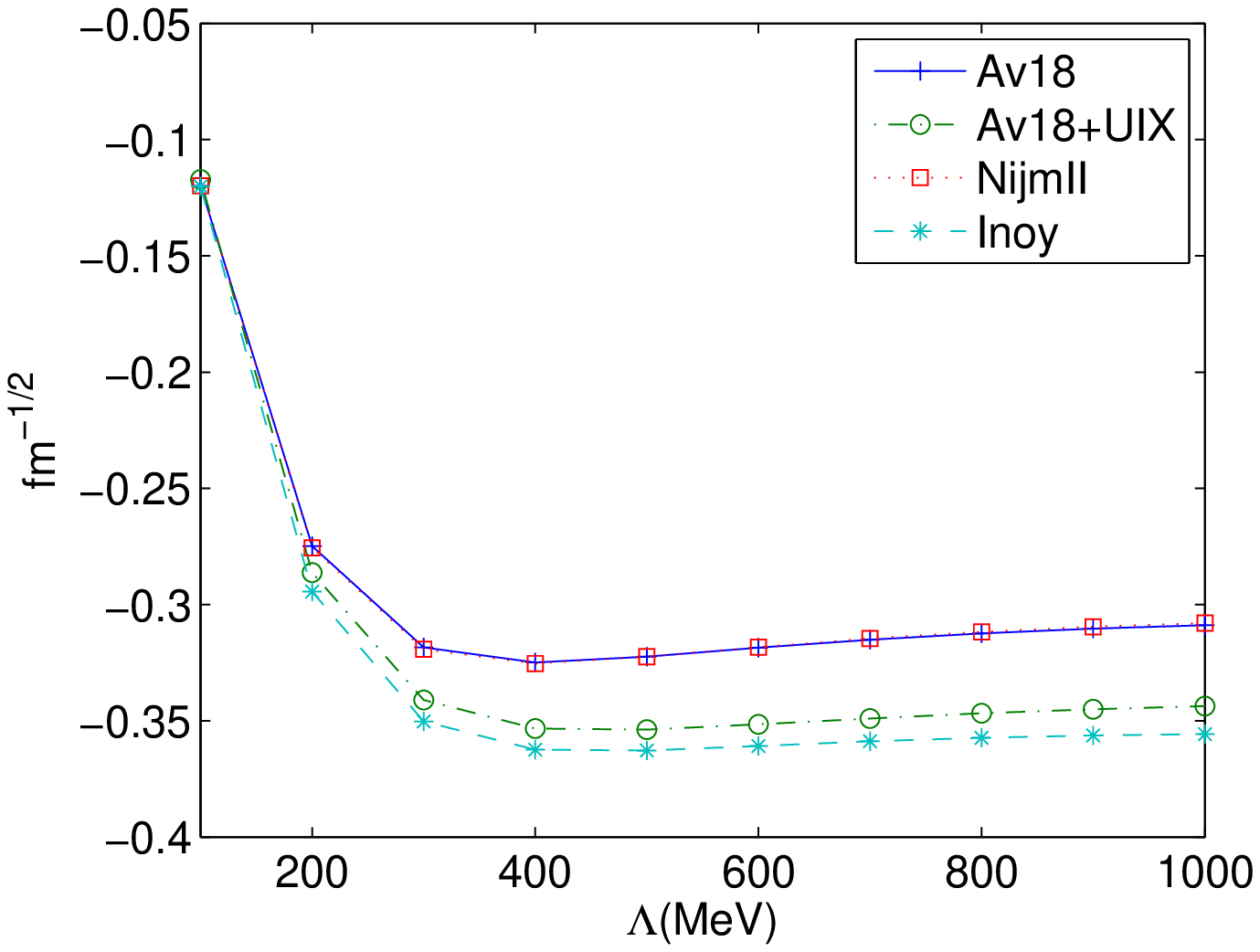}}
\subfigure[$\Lambda^2\widetilde{\cal E}_{\frac{3}{2},(+)}$ for op1]
            {\includegraphics[width=0.4\textwidth]{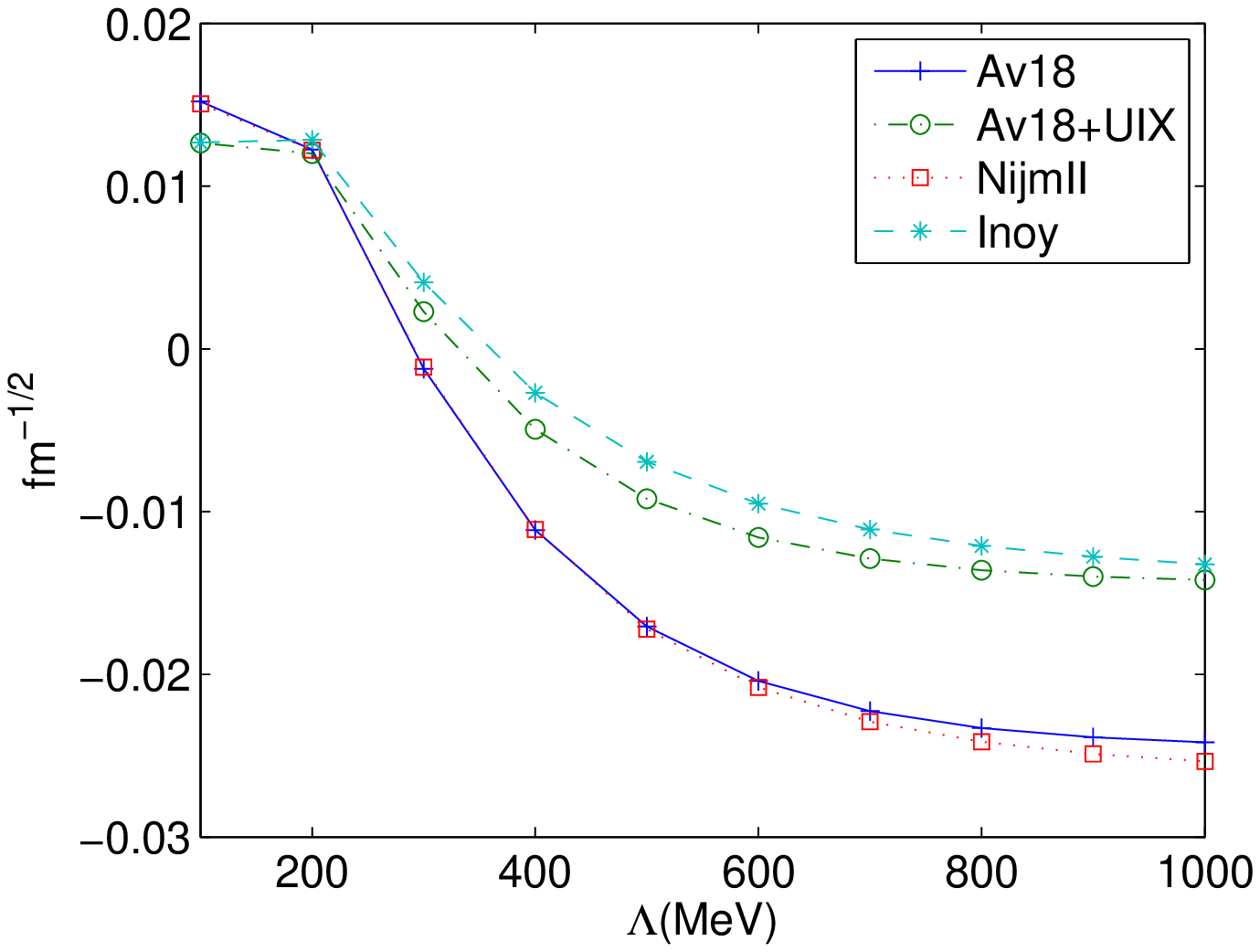}}
\end{center}
    \caption{\label{fig:pionful1:example}
     Cutoff and strong model dependencies of
    $\Lambda^2\widetilde{\cal E}$
    in $fm^{-\frac{1}{2}}$ for $\pi$EFT-I.
    The multiplier $\Lambda^2$ is used to absorb artificial cutoff dependence of $c_n$
    coefficients. }
\end{figure}

To analyze the possible model dependence for the pionful EFT approach, let us consider a contribution of operator 1 to $\Lambda^2\widetilde{\cal E}_{\frac{1}{2},(-)}$
and to $\Lambda^2\widetilde{\cal E}_{\frac{3}{2},(+)}$
 calculated in $\pi$EFT-I approach (see
Fig. \ref{fig:pionful1:example}).
In $\pi$EFT, the physical range for
 cutoff mass scale parameter $\Lambda$ is
about $500< \Lambda < 800$ MeV. One can observe a
rather important dependence on strong potential model in this region.
We cannot  discern long range from short range model dependency
unless all LECs are  determined.
However, a smaller range of the variation of matrix elements for different strong potentials at the pion mass scale indicates that the
contribution of the long range part of strong potentials to the region
of the interest ($500< \Lambda < 800$ MeV) is small.
This means that the large model dependence in
this range ($500< \Lambda < 800$ MeV) is due to short range
part of the wave function, therefore this cutoff
and model dependence
should be absorbed by  higher order contact terms.

Though the general behavior of the matrix elements are consistent with
the expectations of EFT, the 3-body system is rather complicated one to see the direct
relations between the 2-body PV potential and 3-body PV matrix elements.
Therefore, it is useful to re-analyze
the two-body n-p capture process, for which the large model
dependence for a circular polarization of photons, $P^\gamma$,
was reported in \cite{Schiavilla:2004wn}.

\subsection{Two body radiative capture($n+p\to d+\gamma$) }

Parity violating
asymmetry of photons for polarized neutron capture on protons and and their
circular polarization  for the case of unpolarized neutron capture can be written as
\bea
a^\gamma&=&\frac{-\sqrt{2}{\rm Re}\{
           M_1^*(^1S_0)E_1(^3S_1)\}}{|M_1(^1S_0)|^2} \\
P^\gamma       &=&\frac{2{\rm Re}[M_1(^1S_0)E^*_1(^1S_0)]}
        {|M_1(^1S_0)|^2}.
\eea
Here, we neglected
$M_1(^3S_1)$, and $E_1(^1P_1\leftarrow ^3S_1)$ amplitudes.
The $E_1(^3S_1)$ amplitude is a sum of amplitudes with contributions from
parity violating bound state wave function
and from parity violating scattering wave
$(^3P_1\leftarrow ^3S_1)$.
Since $M_1(^3S_1)$ amplitude is suppressed,   one can consider only $E_1(^1S_0)$ contribution
to the $P^\gamma$,  which is dominated by $\rho$ and
 $\omega$ meson exchanges in the DDH formalism.  (The
$a_n^\gamma$ is dominated by one-pion exchange.)

The parity conserving   M1 amplitude can be written as
\bea
M1(^1S_0)=i\frac{\omega \mu_N}{\sqrt{6\pi}\sqrt{4\pi}}\widetilde{\cal M}
         =i\frac{\omega \mu_N}{\sqrt{6\pi}\sqrt{4\pi}}
          \left(\sqrt{4\pi}\sqrt{3}(393.06)\ fm^{\frac{3}{2}}\right).
\eea

Then, PV observables  can be written as
\bea
a_n^\gamma&=&\sum_{m} \left(\frac{c_m}{\mu_N}\right)
           (-\sqrt{8\pi})\frac{\widetilde{\cal E}^{(m)}(^3S_1)}
           {\widetilde{\cal M}(^1S_0)},\no
P^\gamma&=&\sum_{m} \left(\frac{c_m}{\mu_N}\right)
           (-2\sqrt{4\pi})\frac{\widetilde{\cal E}^{(m)}(^1S_0)}
           {\widetilde{\cal M}(^1S_0)}.
\eea
Using strong AV18 and weak DDH-II potentials, one can obtain
\begin{eqnarray}\label{param2}
a^{\gamma}_n &=& 0.15 h^{1}_{\pi}+0.00137h^{1}_{\rho}-
0.00405 h^{1}_{\omega}-0.00137h^{\prime 1}_{\rho}, \\
P^{\gamma} &=& -0.0104 h^{0}_{\rho}-0.00817 h^{2}_{\rho}+0.0111h^{0}_{\omega}.
\end{eqnarray}

\begin{table}[H]%[H] add [H] placement to break table across pages
\caption{\label{tbl:2body:ddh}
Two-body Parity violating observables for potential models
with DDH-best parameter values and Bowman's 4-parameter fits.
}
\begin{ruledtabular}
\begin{tabular}{c rr|rr}
 models    &\multicolumn{2}{c}{${a}^\gamma_n$}&\multicolumn{2}{c}{$P_\gamma$}\\
           & DDH-best & 4-para. fit & DDH-best & 4-para. fit \\
  \hline
AV18 +DDH-I  &$5.25\times 10^{-8} $&$-4.91\times 10^{-9}$&$6.94\times 10^{-9}$&$4.76\times 10^{-9}$\\
AV18 +DDH-II &$5.29\times 10^{-8} $&$-4.81\times 10^{-9}$&$1.76\times 10^{-8}$&$3.01\times 10^{-8}$\\
NijmII+DDH-II&$5.37\times 10^{-8} $&$-4.99\times 10^{-9}$&$2.61\times 10^{-8}$&$6.41\times 10^{-8}$\\
Reid+DDH-II&$5.33\times 10^{-8} $&$-4.85\times 10^{-9}$&$2.65\times 10^{-8}$&$4.68\times 10^{-8}$\\
INOY+DDH-II& $5.60\times 10^{-8} $&$-3.94\times 10^{-9}$&$2.55\times 10^{-7}$&$9.68\times 10^{-7}$\\
\end{tabular}
\end{ruledtabular}
\end{table}

The calculated values of PV observables
for different sets of strong potentials and different choices  of DDH  coupling constants are summarized in Table \ref{tbl:2body:ddh}.
One can see that the circular polarization $P^\gamma$, being dominated by
heavy meson exchange, shows large model dependence in  agreement with the  analysis of n-d case.

\begin{figure}[H]
\begin{center}
        \subfigure[$\mu^2\widetilde{\cal E}_{1,(+)}$ of operator 1]{
            \label{fig:first}
        \includegraphics[width=0.4\textwidth]{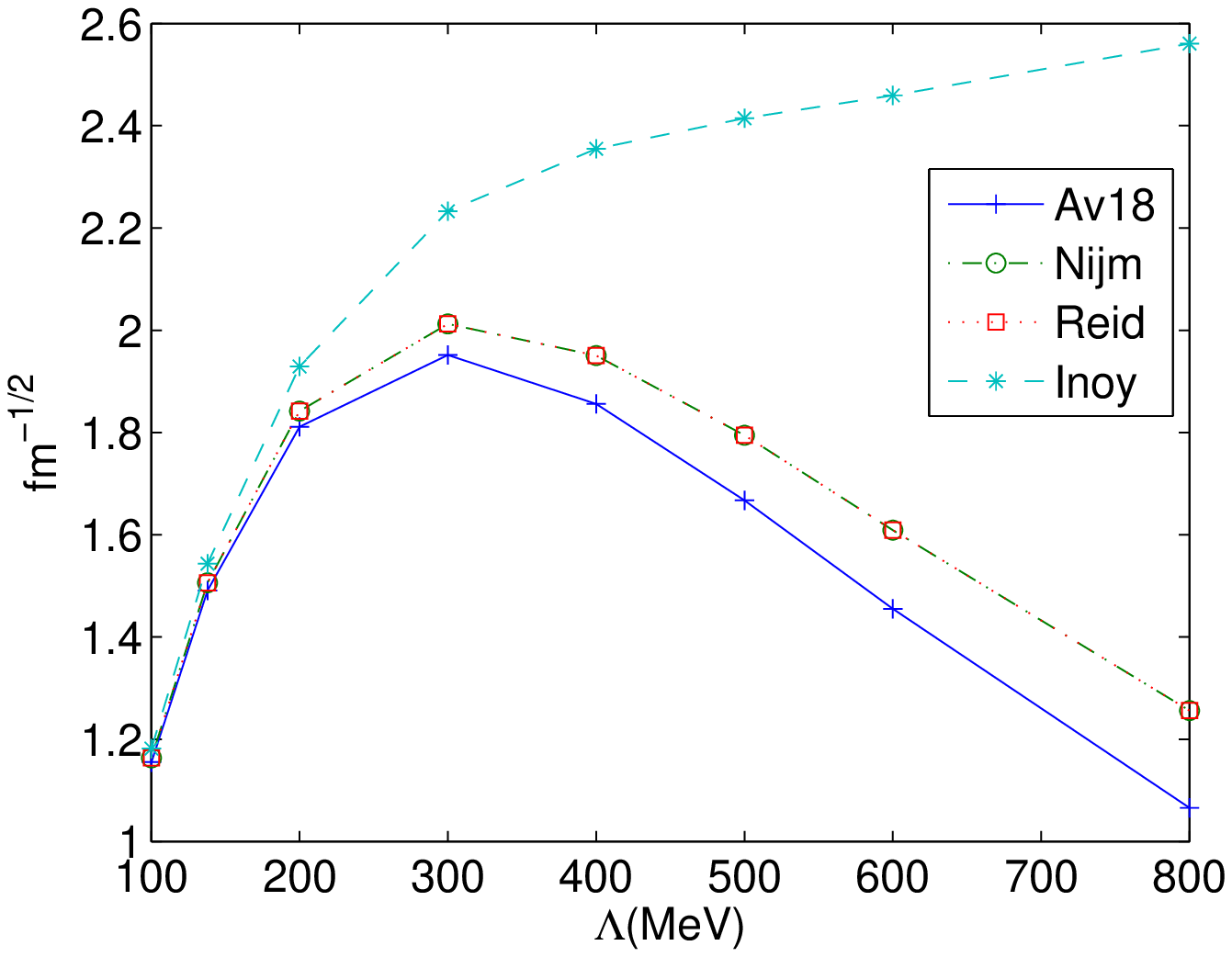}}
        \subfigure[$\mu^2\widetilde{\cal E}_{0,(-)}$ of operator 9]{
           \label{fig:second}
         \includegraphics[width=0.4\textwidth]{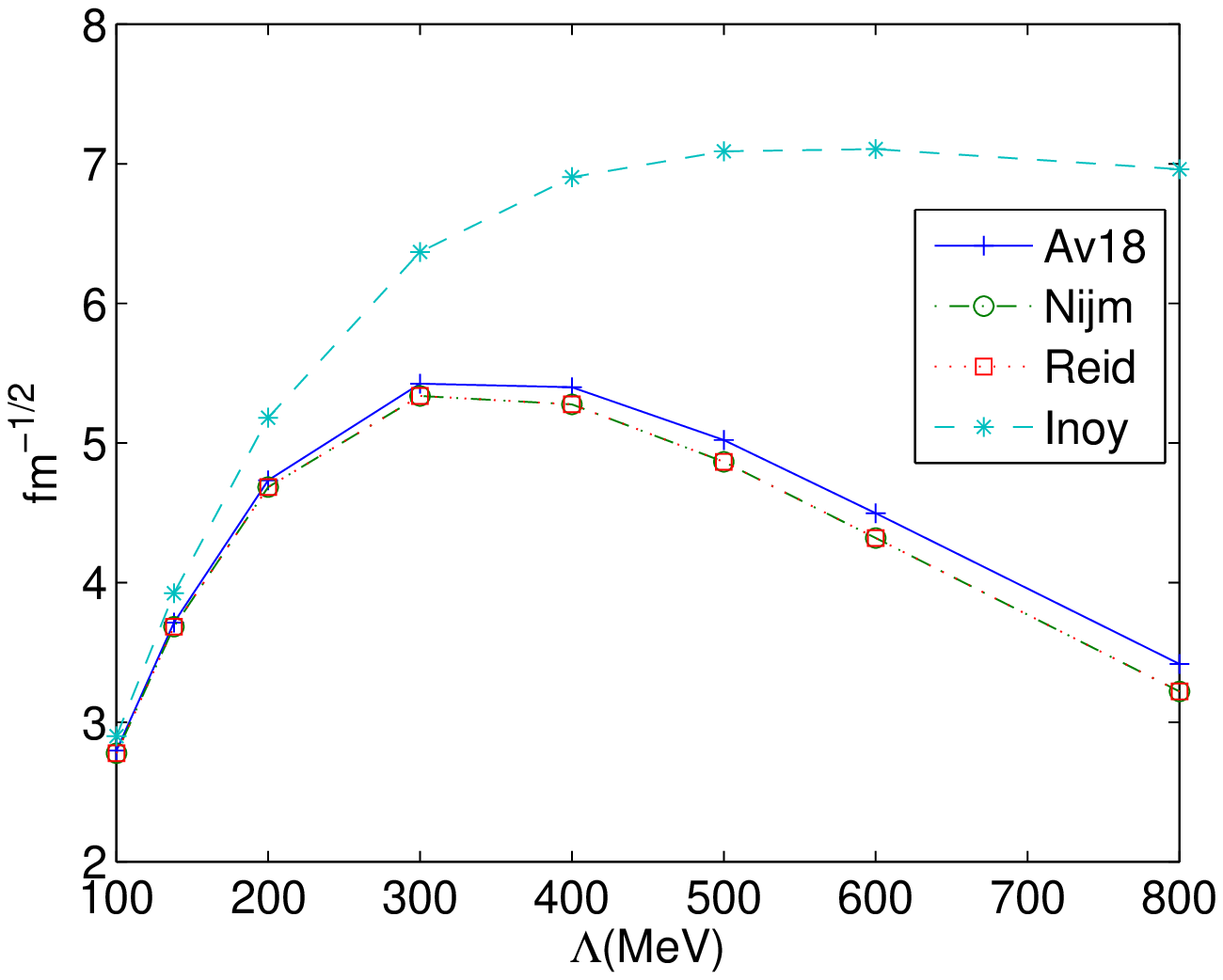}}\\
    \end{center}
    \caption{
    \label{fig:pionless1:2body:example}
    PV transition amplitudes
    $\mu^2\widetilde{\cal E}_{J,(+)}$ in $fm^{-\frac{1}{2}}$
    for AV18 strong potential and PV $\not{\pi}EFT$-I potential.
    }
%\mbox{\epsfxsize=6cm\epsffile{CubicHermite.eps}}
\end{figure}
The cutoff and  model dependence of the transition matrix elements calculated for operators 1 and 9  using $\not{\pi}$EFT-I approach, shown in Figure \ref{fig:pionless1:2body:example},  remind the corresponding
cutoff and model dependencies for the n-d capture process.
One can see that the model dependence is more pronounced
at the scale of  $\rho$ or $\omega$
meson masses in comparison to the pion mass region scale.
This  is also consistent with the statement given for the hybrid approach
that  one can regularize
the short distance contributions by  introducing a cutoff parameter
and, as a consequence,  reduce the uncertainty related to  the short range interactions.
This  indicates that   the possible
reason for model and cutoff dependencies in 3-body (n-d capture) process has the same origin
as those in 2-body case, and it can be treated regularly in the hybrid approach.

For the completeness of the analysis, we present  contributions of different PV operators to PV observables calculated
 in $\not{\pi}$EFT and $\pi$EFT approaches with
the AV18 potential (see
tables \ref{tbl:2body:observables:pionless} and
\ref{tbl:2body:observables:pionful}, correspondingly).
The large difference between matrix elements with pionless and pionful  PV potentials
could be explained by different scales of cutoff parameters, and  the comparison of th
e results obtained from these two approaches   could be done only after the
renormalization of low energy constants.

\begin{table}[H]%[H] add [H] placement to break table across pages
\caption{\label{tbl:2body:observables:pionless}
Two body Parity violating observables for AV18 and $\not{\pi}$EFT potential at $\mu=138$ MeV.
}
\begin{ruledtabular}
\begin{tabular}{c cc|cc}
   &         $\not{\pi}$EFT-I &
   &         $\not{\pi}$EFT-II &                  \\
   \hline
 n & $\mu^2{a}_n^{(n)}$ & $\mu^2{P}^{(n)}_\gamma$
   & $\mu^2{a}_n^{(n)}$ & $\mu^2{P}^{(n)}_\gamma$ \\
 \hline
1 &  $6.02\times 10^{-3}$ &           0
  &  $1.36\times 10^{-2}$ &           0\\
6 &          0            &$2.48\times 10^{-2}$
  & 0& $5.71\times 10^{-2}$\\
8 &          0            &$1.99\times 10^{-2}$
  & 0& $1.78\times 10^{-2}$\\
9 &          0            &$-2.17\times 10^{-2}$
  & 0&$-5.97\times 10^{-2}$\\
\end{tabular}
\end{ruledtabular}
\end{table}

\begin{table}[H]
\caption{\label{tbl:2body:observables:pionful}
Two body Parity violating observables for AV18 and ${\pi}$EFT potential
at $\Lambda=600$ MeV. Only non-vanishing matrix elements are shown.
}
\begin{ruledtabular}
\begin{tabular}{rrr|rrr}
   &   $\pi$EFT-I  & $\pi$EFT-II&
   & $\pi$EFT-I   & $\pi$EFT-II \\
 op&  $a_n^{\gamma(n)}$ & $a_n^{\gamma(n)}$
&op  & $P_\gamma^{(n)}$ & $P_\gamma^{(n)}$\\
\hline
1 &  $ 1.22\times 10^{-2}$ &  $ 7.14\times 10^{-3}$
  & 6 &$ 3.05\times 10^{-3}$ & $  2.01\times 10^{-3}$\\
13& $ 1.11\times 10^{-3} $&$   5.59\times 10^{-4}$
  & 8 &$ 2.70\times 10^{-3}$&$  1.80\times 10^{-3}$\\
14&  $ 1.33\times 10^{-3} $ &$   7.41\times 10^{-4}$
  & 9 &$-4.13\times 10^{-3}$&$ -2.36\times 10^{-3}$\\
15 & $ 2.17\times 10^{-3} $&$   1.00\times 10^{-3}$
  &  &    &\\
\end{tabular}
\end{ruledtabular}
\end{table}

\section{Conclusion}

 PV effects in neutron-deuteron radiative capture are calculated for
DDH-type  and  EFT-type, pionless and pionful, weak interaction potentials.
Three-body problem was solved using Faddeev equations in configuration space,
also by varying the strong interaction part of the Hamiltonian.
Number of different phenomenological  strong potentials has been tested,
including  AV18 NN interaction in conjunction with UIX  3-nucleon force.
The analysis of the obtained results  shows that the values of PV amplitudes
depend both on the choice of the weak as well as strong interaction model.
We demonstrated that this dependence has the expected behavior in the
framework of the standard pionless and pionful EFT approaches.
Therefore, this dependence is expected to be absorbed by LECs.
Nevertheless, in order to obtain model independent  EFT predictions for PV observables,
 one should perform all the calculations in a self-consistent way \cite{Gudkov:2010pt}.
  Using the ``hybrid'' approach we can minimize the model dependence, provided that all
  LECs are defined from the sufficiently large set of  experimental data,
  which does not look practical in the nearest future.

For the case of the DDH approach, the observed model dependence indicates intrinsic difficulty in the description of nuclear PV effects  and could be the reason for the observed discrepancies in the nuclear PV data analysis (see, for example \cite{Holstein:2009zzb} and referencies therein). Thus, the DDH approach could be a reasonable approach for the parametrization and for the analysis of PV effects only if exactly the same strong and weak potentials are used in calculating all PV observables in all nuclei. However, the  existing calculations of nuclear PV effects have been done using different potentials; therefore, strictly speaking, one cannot compare the existing results of these calculations among themselves.
Further, most of the existing calculations do not include
three body interaction which is shown to be important.

We would like to mention that the observed
sensitivity of PV effects to
short range parts of interactions
could be used as a new method for the study of short ranges  nuclear forces.  Once the theory PV effects is well understood, or once we use exactly the same parametrization for weak interactions,
 PV effects can be used to probe short distance dynamics of different nuclear systems described by different strong potentials.

\appendix*


\begin{acknowledgments}
This work was supported by the DOE grants no. DE-FG02-09ER41621.
This work was granted access to the HPC resources of IDRIS
under the allocation 2009-i2009056006
made by GENCI (Grand Equipement National de Calcul Intensif).
We thank the staff members of the IDRIS for their constant help.
\end{acknowledgments}

\bibliography{ParityViolation}

\begin{thebibliography}{38}
\expandafter\ifx\csname natexlab\endcsname\relax\def\natexlab#1{#1}\fi
\expandafter\ifx\csname bibnamefont\endcsname\relax
  \def\bibnamefont#1{#1}\fi
\expandafter\ifx\csname bibfnamefont\endcsname\relax
  \def\bibfnamefont#1{#1}\fi
\expandafter\ifx\csname citenamefont\endcsname\relax
  \def\citenamefont#1{#1}\fi
\expandafter\ifx\csname url\endcsname\relax
  \def\url#1{\texttt{#1}}\fi
\expandafter\ifx\csname urlprefix\endcsname\relax\def\urlprefix{URL }\fi
\providecommand{\bibinfo}[2]{#2}
\providecommand{\eprint}[2][]{\url{#2}}

\bibitem[{\citenamefont{Zhu et~al.}(2005)\citenamefont{Zhu, Maekawa, Holstein,
  Ramsey-Musolf, and van Kolck}}]{Zhu:2004vw}
\bibinfo{author}{\bibfnamefont{S.-L.} \bibnamefont{Zhu}},
  \bibinfo{author}{\bibfnamefont{C.~M.} \bibnamefont{Maekawa}},
  \bibinfo{author}{\bibfnamefont{B.~R.} \bibnamefont{Holstein}},
  \bibinfo{author}{\bibfnamefont{M.~J.} \bibnamefont{Ramsey-Musolf}},
  \bibnamefont{and} \bibinfo{author}{\bibfnamefont{U.}~\bibnamefont{van
  Kolck}}, \bibinfo{journal}{Nucl. Phys.} \textbf{\bibinfo{volume}{A748}},
  \bibinfo{pages}{435} (\bibinfo{year}{2005}).

\bibitem[{\citenamefont{Holstein}(2005{\natexlab{a}})}]{HolsteinUSC}
\bibinfo{author}{\bibfnamefont{B.}~\bibnamefont{Holstein}},
  \emph{\bibinfo{title}{Neutrons and hadronic parity violation}}
  (\bibinfo{year}{2005}{\natexlab{a}}), \bibinfo{note}{proc. of Int. Workshop
  on Theoretical Problems in Fundamental Neutron Physics, October 14-15, 2005,
  Columbia, SC}, \eprint{http://www.physics.sc.edu/TPFNP/Talks/Program.html}.

\bibitem[{\citenamefont{Desplanque}(2005)}]{DesplanqueUSC}
\bibinfo{author}{\bibfnamefont{B.}~\bibnamefont{Desplanque}},
  \emph{\bibinfo{title}{Weak couplings: a few remarks}} (\bibinfo{year}{2005}),
  \bibinfo{note}{proc. of Int. Workshop on Theoretical Problems in Fundamental
  Neutron Physics, October 14-15, 2005, Columbia, SC},
  \eprint{http://www.physics.sc.edu/TPFNP/Talks/Program.html}.

\bibitem[{\citenamefont{Ramsey-Musolf and Page}(2006)}]{RamseyMusolf:2006dz}
\bibinfo{author}{\bibfnamefont{M.~J.} \bibnamefont{Ramsey-Musolf}}
  \bibnamefont{and} \bibinfo{author}{\bibfnamefont{S.~A.} \bibnamefont{Page}},
  \bibinfo{journal}{Ann. Rev. Nucl. Part. Sci.} \textbf{\bibinfo{volume}{56}},
  \bibinfo{pages}{1} (\bibinfo{year}{2006}).

\bibitem[{\citenamefont{Desplanques
  et~al.}(1980{\natexlab{a}})\citenamefont{Desplanques, Donoghue, and
  Holstein}}]{Desplanques1980}
\bibinfo{author}{\bibfnamefont{B.}~\bibnamefont{Desplanques}},
  \bibinfo{author}{\bibfnamefont{J.~F.} \bibnamefont{Donoghue}},
  \bibnamefont{and} \bibinfo{author}{\bibfnamefont{B.~R.}
  \bibnamefont{Holstein}}, \bibinfo{journal}{Annals of Physics}
  \textbf{\bibinfo{volume}{124}}, \bibinfo{pages}{449 }
  (\bibinfo{year}{1980}{\natexlab{a}}).

\bibitem[{\citenamefont{Liu}(2007)}]{Liu:2006dm}
\bibinfo{author}{\bibfnamefont{C.~P.} \bibnamefont{Liu}},
  \bibinfo{journal}{Phys. Rev.} \textbf{\bibinfo{volume}{C75}},
  \bibinfo{pages}{065501} (\bibinfo{year}{2007}).

\bibitem[{\citenamefont{Girlanda}(2008)}]{Girlanda:2008ts}
\bibinfo{author}{\bibfnamefont{L.}~\bibnamefont{Girlanda}},
  \bibinfo{journal}{Phys. Rev.} \textbf{\bibinfo{volume}{C77}},
  \bibinfo{pages}{067001} (\bibinfo{year}{2008}), \eprint{0804.0772}.

\bibitem[{\citenamefont{Phillips et~al.}(2009)\citenamefont{Phillips,
  Schindler, and Springer}}]{Phillips:2008hn}
\bibinfo{author}{\bibfnamefont{D.~R.} \bibnamefont{Phillips}},
  \bibinfo{author}{\bibfnamefont{M.~R.} \bibnamefont{Schindler}},
  \bibnamefont{and} \bibinfo{author}{\bibfnamefont{R.~P.}
  \bibnamefont{Springer}}, \bibinfo{journal}{Nucl. Phys.}
  \textbf{\bibinfo{volume}{A822}}, \bibinfo{pages}{1} (\bibinfo{year}{2009}).

\bibitem[{\citenamefont{Shin et~al.}(2010)\citenamefont{Shin, Ando, and
  Hyun}}]{Shin:2009hi}
\bibinfo{author}{\bibfnamefont{J.~W.} \bibnamefont{Shin}},
  \bibinfo{author}{\bibfnamefont{S.}~\bibnamefont{Ando}}, \bibnamefont{and}
  \bibinfo{author}{\bibfnamefont{C.~H.} \bibnamefont{Hyun}},
  \bibinfo{journal}{Phys. Rev.} \textbf{\bibinfo{volume}{C81}},
  \bibinfo{pages}{055501} (\bibinfo{year}{2010}).

\bibitem[{\citenamefont{Schindler and Springer}(2009)}]{Schindler:2009wd}
\bibinfo{author}{\bibfnamefont{M.~R.} \bibnamefont{Schindler}}
  \bibnamefont{and} \bibinfo{author}{\bibfnamefont{R.~P.}
  \bibnamefont{Springer}} (\bibinfo{year}{2009}), \eprint{0907.5358}.

\bibitem[{\citenamefont{Sushkov and Flambaum}(1982)}]{Sushkov:1982fa}
\bibinfo{author}{\bibfnamefont{O.~P.} \bibnamefont{Sushkov}} \bibnamefont{and}
  \bibinfo{author}{\bibfnamefont{V.~V.} \bibnamefont{Flambaum}},
  \bibinfo{journal}{Sov. Phys. Usp.} \textbf{\bibinfo{volume}{25}},
  \bibinfo{pages}{1} (\bibinfo{year}{1982}).

\bibitem[{\citenamefont{Bunakov and Gudkov}(1983)}]{Bunakov:1982is}
\bibinfo{author}{\bibfnamefont{V.~E.} \bibnamefont{Bunakov}} \bibnamefont{and}
  \bibinfo{author}{\bibfnamefont{V.~P.} \bibnamefont{Gudkov}},
  \bibinfo{journal}{Nucl. Phys.} \textbf{\bibinfo{volume}{A401}},
  \bibinfo{pages}{93} (\bibinfo{year}{1983}).

\bibitem[{\citenamefont{Gudkov}(1992)}]{Gudkov:1991qg}
\bibinfo{author}{\bibfnamefont{V.~P.} \bibnamefont{Gudkov}},
  \bibinfo{journal}{Phys. Rept.} \textbf{\bibinfo{volume}{212}},
  \bibinfo{pages}{77} (\bibinfo{year}{1992}).

\bibitem[{\citenamefont{Schiavilla et~al.}(2008)\citenamefont{Schiavilla,
  Viviani, Girlanda, Kievsky, and Marcucci}}]{Schiavilla:2008ic}
\bibinfo{author}{\bibfnamefont{R.}~\bibnamefont{Schiavilla}},
  \bibinfo{author}{\bibfnamefont{M.}~\bibnamefont{Viviani}},
  \bibinfo{author}{\bibfnamefont{L.}~\bibnamefont{Girlanda}},
  \bibinfo{author}{\bibfnamefont{A.}~\bibnamefont{Kievsky}}, \bibnamefont{and}
  \bibinfo{author}{\bibfnamefont{L.~E.} \bibnamefont{Marcucci}},
  \bibinfo{journal}{Phys. Rev.} \textbf{\bibinfo{volume}{C78}},
  \bibinfo{pages}{014002} (\bibinfo{year}{2008}).

\bibitem[{\citenamefont{Song et~al.}(2011{\natexlab{a}})\citenamefont{Song,
  Lazauskas, and Gudkov}}]{Song:2010sz}
\bibinfo{author}{\bibfnamefont{Y.-H.} \bibnamefont{Song}},
  \bibinfo{author}{\bibfnamefont{R.}~\bibnamefont{Lazauskas}},
  \bibnamefont{and} \bibinfo{author}{\bibfnamefont{V.}~\bibnamefont{Gudkov}},
  \bibinfo{journal}{Phys.Rev.} \textbf{\bibinfo{volume}{C83}},
  \bibinfo{pages}{015501} (\bibinfo{year}{2011}{\natexlab{a}}),
  \eprint{1011.2221}.

\bibitem[{\citenamefont{Moskalev}(1969)}]{Moskalev:1969}
\bibinfo{author}{\bibfnamefont{A.}~\bibnamefont{Moskalev}},
  \bibinfo{journal}{Sov. J. of Nucl. Phys.} \textbf{\bibinfo{volume}{9}},
  \bibinfo{pages}{99} (\bibinfo{year}{1969}).

\bibitem[{\citenamefont{E.~Hadjimichael and Newton}(1974)}]{Hadjimichael:1974}
\bibinfo{author}{\bibfnamefont{E.~H.} \bibnamefont{E.~Hadjimichael}}
  \bibnamefont{and} \bibinfo{author}{\bibfnamefont{V.}~\bibnamefont{Newton}},
  \bibinfo{journal}{Nucl. Phys.} \textbf{\bibinfo{volume}{A228}},
  \bibinfo{pages}{1} (\bibinfo{year}{1974}).

\bibitem[{\citenamefont{McKellar}(1974)}]{McKellar:1974yr}
\bibinfo{author}{\bibfnamefont{B.~H.} \bibnamefont{McKellar}},
  \bibinfo{journal}{Phys.Rev.} \textbf{\bibinfo{volume}{C9}},
  \bibinfo{pages}{1790} (\bibinfo{year}{1974}).

\bibitem[{\citenamefont{Desplanques and Benayoun}(1986)}]{Desplanques:1986cq}
\bibinfo{author}{\bibfnamefont{B.}~\bibnamefont{Desplanques}} \bibnamefont{and}
  \bibinfo{author}{\bibfnamefont{J.}~\bibnamefont{Benayoun}},
  \bibinfo{journal}{Nucl.Phys.} \textbf{\bibinfo{volume}{A458}},
  \bibinfo{pages}{689} (\bibinfo{year}{1986}).

\bibitem[{\citenamefont{Song et~al.}(2009)\citenamefont{Song, Lazauskas, and
  Park}}]{Song:2008zf}
\bibinfo{author}{\bibfnamefont{Y.-H.} \bibnamefont{Song}},
  \bibinfo{author}{\bibfnamefont{R.}~\bibnamefont{Lazauskas}},
  \bibnamefont{and} \bibinfo{author}{\bibfnamefont{T.-S.} \bibnamefont{Park}},
  \bibinfo{journal}{Phys. Rev.} \textbf{\bibinfo{volume}{C79}},
  \bibinfo{pages}{064002} (\bibinfo{year}{2009}).

\bibitem[{\citenamefont{Pastore et~al.}(2009)\citenamefont{Pastore, Girlanda,
  Schiavilla, Viviani, and Wiringa}}]{Pastore:2009is}
\bibinfo{author}{\bibfnamefont{S.}~\bibnamefont{Pastore}},
  \bibinfo{author}{\bibfnamefont{L.}~\bibnamefont{Girlanda}},
  \bibinfo{author}{\bibfnamefont{R.}~\bibnamefont{Schiavilla}},
  \bibinfo{author}{\bibfnamefont{M.}~\bibnamefont{Viviani}}, \bibnamefont{and}
  \bibinfo{author}{\bibfnamefont{R.~B.} \bibnamefont{Wiringa}},
  \bibinfo{journal}{Phys. Rev.} \textbf{\bibinfo{volume}{C80}},
  \bibinfo{pages}{034004} (\bibinfo{year}{2009}).

\bibitem[{\citenamefont{Song et~al.}(2007)\citenamefont{Song, Lazauskas, Park,
  and Min}}]{Song:2007bj}
\bibinfo{author}{\bibfnamefont{Y.-H.} \bibnamefont{Song}},
  \bibinfo{author}{\bibfnamefont{R.}~\bibnamefont{Lazauskas}},
  \bibinfo{author}{\bibfnamefont{T.-S.} \bibnamefont{Park}}, \bibnamefont{and}
  \bibinfo{author}{\bibfnamefont{D.-P.} \bibnamefont{Min}},
  \bibinfo{journal}{Phys. Lett.} \textbf{\bibinfo{volume}{B656}},
  \bibinfo{pages}{174} (\bibinfo{year}{2007}).

\bibitem[{\citenamefont{Lazauskas et~al.}(2009)\citenamefont{Lazauskas, Song,
  and Park}}]{Lazauskas:2009nw}
\bibinfo{author}{\bibfnamefont{R.}~\bibnamefont{Lazauskas}},
  \bibinfo{author}{\bibfnamefont{Y.-H.} \bibnamefont{Song}}, \bibnamefont{and}
  \bibinfo{author}{\bibfnamefont{T.-S.} \bibnamefont{Park}}
  (\bibinfo{year}{2009}), \eprint{0905.3119}.

\bibitem[{\citenamefont{Park et~al.}(2003)}]{Park:2002yp}
\bibinfo{author}{\bibfnamefont{T.~S.} \bibnamefont{Park}} \bibnamefont{et~al.},
  \bibinfo{journal}{Phys. Rev.} \textbf{\bibinfo{volume}{C67}},
  \bibinfo{pages}{055206} (\bibinfo{year}{2003}).

\bibitem[{\citenamefont{Girlanda et~al.}(2010)}]{Girlanda:2010vm}
\bibinfo{author}{\bibfnamefont{L.}~\bibnamefont{Girlanda}} \bibnamefont{et~al.}
  (\bibinfo{year}{2010}), \eprint{1008.0356}.

\bibitem[{\citenamefont{Song et~al.}(2011{\natexlab{b}})\citenamefont{Song,
  Lazauskas, and Gudkov}}]{Song:2011sw}
\bibinfo{author}{\bibfnamefont{Y.-H.} \bibnamefont{Song}},
  \bibinfo{author}{\bibfnamefont{R.}~\bibnamefont{Lazauskas}},
  \bibnamefont{and} \bibinfo{author}{\bibfnamefont{V.}~\bibnamefont{Gudkov}},
  \bibinfo{journal}{Phys.Rev.} \textbf{\bibinfo{volume}{C83}},
  \bibinfo{pages}{065503} (\bibinfo{year}{2011}{\natexlab{b}}),
  \eprint{1104.3051}.

\bibitem[{\citenamefont{Song et~al.}(2011{\natexlab{c}})\citenamefont{Song,
  Lazauskas, and Gudkov}}]{Song:2011jh}
\bibinfo{author}{\bibfnamefont{Y.-H.} \bibnamefont{Song}},
  \bibinfo{author}{\bibfnamefont{R.}~\bibnamefont{Lazauskas}},
  \bibnamefont{and} \bibinfo{author}{\bibfnamefont{V.}~\bibnamefont{Gudkov}},
  \bibinfo{journal}{Phys.Rev.} \textbf{\bibinfo{volume}{C84}},
  \bibinfo{pages}{025501} (\bibinfo{year}{2011}{\natexlab{c}}),
  \eprint{1105.1327}.

\bibitem[{\citenamefont{Griesshammer et~al.}(2012)\citenamefont{Griesshammer,
  Schindler, and Springer}}]{Griesshammer:2011md}
\bibinfo{author}{\bibfnamefont{H.~W.} \bibnamefont{Griesshammer}},
  \bibinfo{author}{\bibfnamefont{M.~R.} \bibnamefont{Schindler}},
  \bibnamefont{and} \bibinfo{author}{\bibfnamefont{R.~P.}
  \bibnamefont{Springer}}, \bibinfo{journal}{Eur.Phys.J.}
  \textbf{\bibinfo{volume}{A48}}, \bibinfo{pages}{7} (\bibinfo{year}{2012}),
  \eprint{1109.5667}.

\bibitem[{\citenamefont{Desplanques
  et~al.}(1980{\natexlab{b}})\citenamefont{Desplanques, Donoghue, and
  Holstein}}]{Desplanques:1979hn}
\bibinfo{author}{\bibfnamefont{B.}~\bibnamefont{Desplanques}},
  \bibinfo{author}{\bibfnamefont{J.~F.} \bibnamefont{Donoghue}},
  \bibnamefont{and} \bibinfo{author}{\bibfnamefont{B.~R.}
  \bibnamefont{Holstein}}, \bibinfo{journal}{Ann. Phys.}
  \textbf{\bibinfo{volume}{124}}, \bibinfo{pages}{449}
  (\bibinfo{year}{1980}{\natexlab{b}}), ISSN \bibinfo{issn}{0003-4916}.

\bibitem[{\citenamefont{Schiavilla et~al.}(2004)\citenamefont{Schiavilla,
  Carlson, and Paris}}]{Schiavilla:2004wn}
\bibinfo{author}{\bibfnamefont{R.}~\bibnamefont{Schiavilla}},
  \bibinfo{author}{\bibfnamefont{J.}~\bibnamefont{Carlson}}, \bibnamefont{and}
  \bibinfo{author}{\bibfnamefont{M.~W.} \bibnamefont{Paris}},
  \bibinfo{journal}{Phys. Rev.} \textbf{\bibinfo{volume}{C70}},
  \bibinfo{pages}{044007} (\bibinfo{year}{2004}).

\bibitem[{\citenamefont{Hyun et~al.}(2007)\citenamefont{Hyun, Ando, and
  Desplanques}}]{Hyun:2006mp}
\bibinfo{author}{\bibfnamefont{C.}~\bibnamefont{Hyun}},
  \bibinfo{author}{\bibfnamefont{S.}~\bibnamefont{Ando}}, \bibnamefont{and}
  \bibinfo{author}{\bibfnamefont{B.}~\bibnamefont{Desplanques}},
  \bibinfo{journal}{Eur.Phys.J.} \textbf{\bibinfo{volume}{A32}},
  \bibinfo{pages}{513} (\bibinfo{year}{2007}), \eprint{nucl-th/0609015}.

\bibitem[{\citenamefont{Faddeev}(1961)}]{Faddeev:1960su}
\bibinfo{author}{\bibfnamefont{L.~D.} \bibnamefont{Faddeev}},
  \bibinfo{journal}{Sov. Phys. JETP} \textbf{\bibinfo{volume}{12}},
  \bibinfo{pages}{1014} (\bibinfo{year}{1961}).

\bibitem[{\citenamefont{Lazauskas}(2003)}]{These_Rimas_03}
\bibinfo{author}{\bibfnamefont{R.}~\bibnamefont{Lazauskas}}
  (\bibinfo{year}{2003}), \bibinfo{note}{universite Joseph Fourier, Grenoblex}.

\bibitem[{\citenamefont{Lazauskas}(2009)}]{Lazauskas_2008}
\bibinfo{author}{\bibfnamefont{R.}~\bibnamefont{Lazauskas}},
  \bibinfo{journal}{Few-Body Systems} \textbf{\bibinfo{volume}{46}},
  \bibinfo{pages}{37} (\bibinfo{year}{2009}),
  \urlprefix\url{http://arxiv.org/abs/0808.1650}.

\bibitem[{\citenamefont{Bowman}()}]{Bowman}
\bibinfo{author}{\bibfnamefont{J.~D.} \bibnamefont{Bowman}},
  \bibinfo{note}{``Hadronic Weak Interaction'', INT Workshop on Electric Dipole
  Moments and CP Violations, March 19-23, 2007},
  \eprint{http://www.int.washington.edu/talks/WorkShops/$int\_07\_1$/}.

\bibitem[{\citenamefont{Holstein}(2005{\natexlab{b}})}]{Holstein:2006bv}
\bibinfo{author}{\bibfnamefont{B.~R.} \bibnamefont{Holstein}},
  \bibinfo{journal}{Fizika.} \textbf{\bibinfo{volume}{B14}},
  \bibinfo{pages}{165} (\bibinfo{year}{2005}{\natexlab{b}}),
  \eprint{nucl-th/0607038}.

\bibitem[{\citenamefont{Gudkov and Song}(2010)}]{Gudkov:2010pt}
\bibinfo{author}{\bibfnamefont{V.}~\bibnamefont{Gudkov}} \bibnamefont{and}
  \bibinfo{author}{\bibfnamefont{Y.-H.} \bibnamefont{Song}},
  \bibinfo{journal}{Phys. Rev.} \textbf{\bibinfo{volume}{C82}},
  \bibinfo{pages}{028502} (\bibinfo{year}{2010}).

\bibitem[{\citenamefont{Holstein}(2009)}]{Holstein:2009zzb}
\bibinfo{author}{\bibfnamefont{B.~R.} \bibnamefont{Holstein}},
  \bibinfo{journal}{Eur. Phys. J.} \textbf{\bibinfo{volume}{A41}},
  \bibinfo{pages}{279} (\bibinfo{year}{2009}).

\end{thebibliography}

\end{document}